\documentclass[%
 reprint,
 amsmath,amssymb,
 aps,
]{revtex4-1}

\usepackage{graphicx}
\usepackage{dcolumn}
\usepackage{bm}

\usepackage{hyperref}

\begin{document}

\preprint{APS/123-QED}

\title{Skyrmion on magnetic tunnel junction: weaving quantum transport with micro-magnetism}

\author{Aashish Chahal$^1$}
\author{Abhishek Sharma$^{2}$}%
\email{abhi@iitmandi.ac.in}
\affiliation{%
 $^1$ Department Electrical Engineering, Indian Institute of Technology Ropar (Punjab, India)
}%
\affiliation{%
 $^2$ School of Computing and Electrical Engineering, Indian Institute of Technology Mandi (Himachal Pradesh, India)
}%

\date{\today}

\begin{abstract}
Over the last two decades, non-trivial magnetic textures, especially the magnetic skyrmion family, have been extensively explored out of fundamental interest and for diverse possible applications. Given the possible technological and scientific ramifications of skyrmion-texture on magnetic tunnel junctions (Sk-MTJs), in this work, we present a non-equilibrium Green's function (NEGF) based description of Sk-MTJs for both Néel and Bloch textures to capture the spin/charge current across different voltages, temperatures, and sizes. We predict the emergence of a textured spin current from the uniform layer of the Sk-MTJs, along with a radially varying, asymmetric voltage dependence of spin torque. We discuss the identification of Néel-type and Bloch-type skyrmions, including their helicity, based on local measurements of the spin current. We describe the tunneling magnetoresistance (TMR) roll-off in Sk-MTJs with lower cross-sectional areas and higher temperatures based on transmission spectra analysis. Additionally, considering the significant implications of MTJ scaling, we uncover the scaling effects in all skyrmion(AS)-MTJs, where both contacts host skyrmions. We demonstrate the impact of scaling on TMR, spin, and charge currents, providing a pathway for the Sk/AS-MTJs design optimization with miniaturization. We also introduce a computationally efficient, analytically grounded coupled spatio-eigen framework of NEGF, alleviating the sine qua non of the 3D-NEGF for systems that lack translational invariance in the transverse direction, such as Sk/AS-MTJs. 
 
\end{abstract}

\maketitle

\section{\label{sec:level1}Introduction}
The rise of magnetic textures—such as merons, skyrmions, bimerons, antiskyrmions, skyrmioniums, and hopfions—at the nanoscale has opened new avenues for both fundamental exploration and a wide range of practical applications \cite{Beyond}. The magnetic skyrmions \cite{FM1,FM2,FM3,TF1,TF2}, particularly notable for their stable vortex-like spin textures and non-trivial topology \cite{SkyT}, have garnered significant attention for their technological relevance. These micromagnetic structures manifest particle-like attributes due to their topological protection and nanometer-scale dimensions \cite{SkyS}. The combination of these characteristics, coupled with a low depinning current density \cite{SkyI} and the capability for electrical manipulation and detection \cite{SkyMD}, positions magnetic skyrmions as highly auspicious candidates for the development of next-generation memory and processing devices. Consequently, many skyrmion-based applications have been proposed, such as nano-oscillators \cite{SkyN}, racetrack memories \cite{SkyM}, logic gates \cite{SkyL1, SkyL2}, neuromorphic computing \cite{SkyNM1,SkyNM2}, and transistors \cite{SkyTr}. Recent works \cite{SkyQ1,SkyQ2,SkyQ3} based on the quantization of the helicity have delineated the potential of skyrmions, merons \cite{MeronQ} and domain-walls\cite{DomainQ} for quantum computing.\\
\indent From a technological perspective, the electrical detection of skyrmions and other magnetic textures via tunnel magnetoresistance (TMR) in magnetic tunnel junctions (MTJs) holds significant importance. The successful demonstration of nucleation and electrical detection of magnetic skyrmions on MTJs at room temperature \cite{MTS1, MTS2, MTS3, MTS4, MTJ7} represents a key milestone in unlocking the potential of skyrmions. The associated technological implications of single skyrmion (Sk-) on MTJ necessitate a comprehensive non-equilibrium understanding of TMR, along with the charge and spin-current profiles of Sk-MTJs for the aforementioned application designs. The recently proposed scheme to control Qubit operations of skyrmion using spin currents \cite{SkyQ3}, possible electrical read-out of skyrmion, along with the push for skyrmion and MTJ miniaturization for high packing density\cite{Skyy}, further underscores the necessity of understanding the non-equilibrium characteristics of Sk-MTJs. \\
\indent Conventionally, in a regular uniform-textured MTJs, as illustrated in Fig.~\ref{fig:skyrmion}(a) \cite{SC1, SC2}, non-equilibrium Green's function (NEGF) based quantum transport \cite{NEGF}, have theoretically predicted the non-trivial voltage dependence of TMR, charge, and spin currents. These predictions have laid the groundwork for subsequent experimental validations \cite{SC3, SC4}. The non-equilibrium characteristics of regular MTJs, when coupled with the Landau-Lifshitz-Gilbert-Slonczewski (LLGS) equation \cite{LLG1}, provide a robust framework for understanding the dynamic interplay between charge/spin transport and the magnetization dynamics of nanomagnets \cite{NLLG, NS_2,Cirkit}. This framework has also played a pivotal role in shaping the design of various MTJ-based applications \cite{NS_1, NS_2, RAM, M_sen_1, M_sen_2}.\\
\indent However, for devices with varying contact properties along the transverse direction, such as MTJs with micromagnetic contacts (e.g., Sk-MTJs, All-skyrmion (AS)-MTJ) a computationally intensive, full 3-dimensional (3D) NEGF formalism is required. The necessity of 3D-NEGF stems from the lack of translational symmetry and the absence of a common eigenbasis between the contacts and channels along the transverse direction of transport. To address this challenge, we introduce a coupled `spatio-eigen' approach to NEGF, which circumvents the need for full 3D-NEGF solutions. We organize the remainder of this work as follows: Sec.~II, along with analytical underpinnings, demonstrates that the proposed spatio-eigen approach provides an equivalent description of current flow (both spin and charge) while significantly reducing matrix size and computational demand without compromising the underlying physics. Notably, the spatio-eigen NEGF approach is not limited to MTJ-like devices with micromagnetic contacts but is broadly applicable, alleviating the need for a full 3D-NEGF solution. Section~III outlines the modeling parameters used, and in Sec.~IV, using the spatio-eigen NEGF approach, we present a non-equilibrium analysis of Sk-MTJs and AS-MTJs, examining their spin and current profiles as well as TMR across various voltages, temperatures, and device sizes. Finally, in Sec.~V, we conclude with future perspectives on analyzing a broader family of micromagnetic textures on MTJs and potential extensions to combine micromagnetic dynamics with NEGF. We also briefly discuss the feasibility of applying the spatio-eigen approach for efficient self-consistent NEGF and density functional theory (DFT) calculations\cite{DFT1}.
\section{\label{sec:level1} Methodology}
\begin{figure}[htbp]
\centering
\includegraphics[width=\columnwidth]{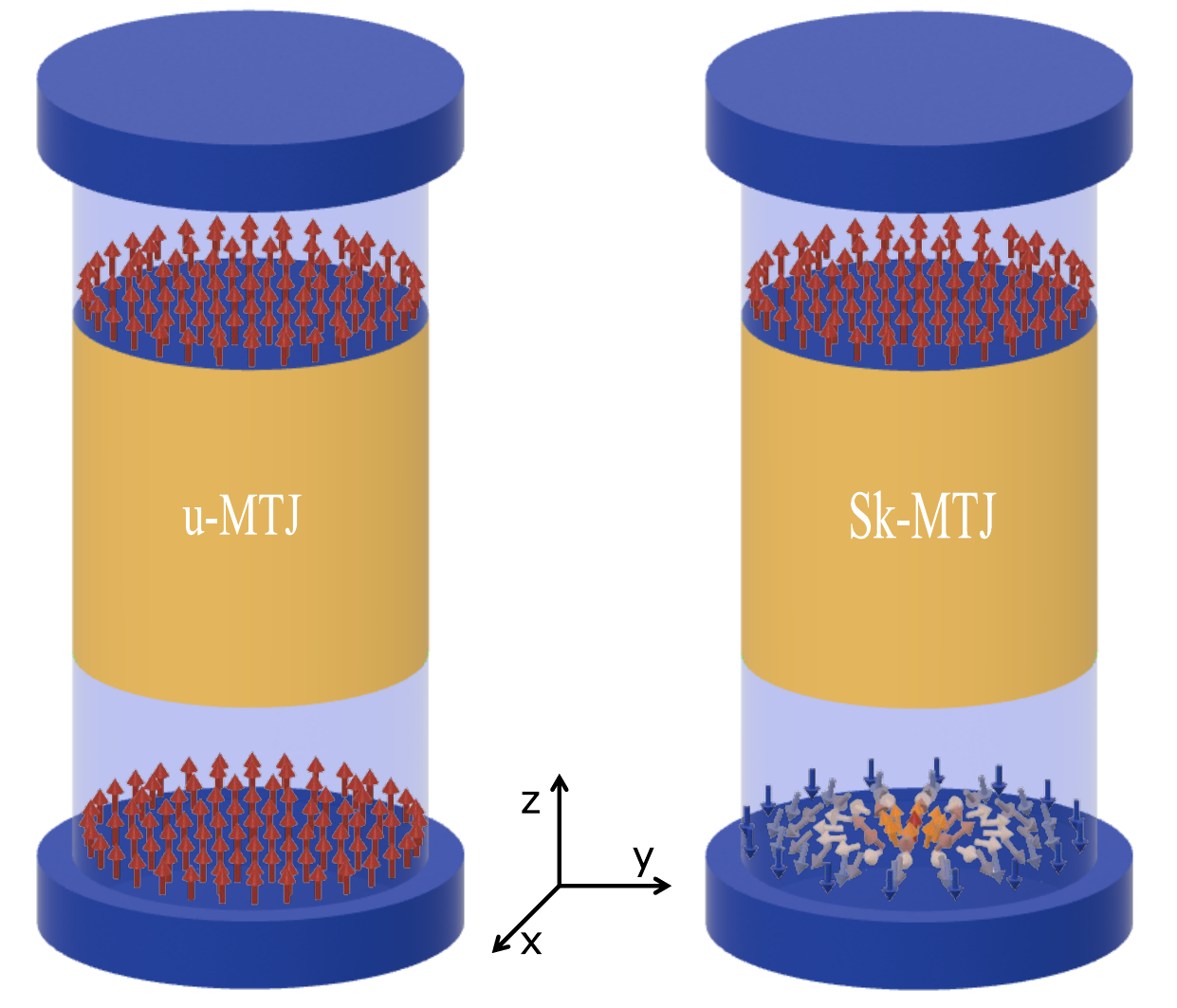}
\caption{Schematics of (a) u-MTJ with uniform magnetization of top/bottom FMs and (b) Sk-MTJ hosting skyrmion in bottom FM.}
\label{fig:skyrmion}
\end{figure}

In this section, we detail NEGF-based quantum transport for the devices having micro-magnetic textured contacts. We present two devices: the regular magnetic tunnel junction (u-MTJ) device, consisting of an insulating channel (MgO)\cite{MTJ_MGO} sandwiched between two ferromagnetic (FM) layers with spatially uniform magnetization texture, and the skyrmion textured magnetic tunnel junction (Sk-MTJ), in which one FM layer exhibits uniform magnetization, and the other layer features a magnetic skyrmion, as depicted in Fig.~\ref{fig:skyrmion}. The u-MTJ is presented for reference purposes. The methodology described subsequently is generic in nature, and extendable to multi-contacts hosting micro-magnetic textures. \\
\indent The model tight-binding Hamiltonian for the system shown in Fig. \ref{fig:skyrmion} is given by
\begin{equation}
H = \sum_i c_i^\dagger \epsilon_i c_i + \sum_{\langle i,j \rangle} c_i^\dagger t_0 c_j - \delta \sum_i c_i^\dagger \boldsymbol{\sigma}_i \cdot \mathbf{m
}_i c_i + \sum_i V_i c_i^\dagger  c_i,
\label{eq:SecQ}
\end{equation}
\noindent here $i, j$ are the site indices, and $\langle i,j \rangle$ indicates a sum over all nearest neighbors, $c_i = [c_{i,\uparrow} \, c_{i,\downarrow}]^\text{T}$ is the annihilation operator, $\mathbf{m}_i$ is the localized spin, $\boldsymbol{\sigma}_i$ is the electron spin and $\epsilon_i$ is an on-site energy term at site $i$, $\delta$ is the spin coupling, $t_0$ is the nearest neighbor hopping term and $V_i$ is potential energy term, including bias voltage and barrier potential (e.g. MgO). We partition the device Hamiltonian $H = H_0 + \textbf{V}$ in two parts, with $H_0$ containing the kinetic and Stoner parts\cite{SC1}, and \textbf{V} containing potential terms (bias voltage and barrier potential). \\
\indent Typically, in MTJs with uniform-magnetic textured contacts (Fig.~\ref{fig:skyrmion}(a)), the translational symmetry in the transverse direction ($\hat{x}$ $\&$ $\hat{y}$) is typically exploited using periodic boundary conditions and Bloch expansion \cite{Ashcroft, Nick}, to manage the computational demand of the NEGF method. However, the lack of translational symmetry in MTJs with micromagnetic textured contacts (such as Fig.~\ref{fig:skyrmion}(b)) breaks the notion of transverse crystal momentum as a valid quantum number in the transverse direction, rendering the Bloch expansion method inapplicable. Consequently, to evaluate the non-equilibrium current-voltage characteristics of MTJs with micromagnetic textures, or any device where contact properties vary along the transverse direction, the use of the computationally intensive three-dimensional (3D) NEGF method becomes inevitable. This computational burden significantly limits the applicability and broader adoption of the NEGF approach for device modeling.\\
\indent Recognizing the heavy computational requirements of 3D-NEGF for the devices where contact properties vary along the transverse direction, such as MTJs with micromagnetic textures, we propose a coupled spatio-eigen framework of NEGF. This approach exploits the finite bandwidth of incoming electron excitation to significantly reduce the computational complexity while maintaining the same physical description of the system. The spatio-eigen formulation of NEGF is indifferent to the translational symmetry along the transverse direction. This methodology reshapes the device Hamiltonian in spatial basis (e.g., discretized tight-binding or linear combination of atomic orbitals \cite{Ashcroft}) along the transport direction (e.g., $\hat{z}$ in Fig.~\ref{fig:skyrmion}) and in the eigenbasis along the transverse direction. \\
\indent To establish the spatio-eigen NEGF formalism, we utilize a finite difference discretized tight-binding model for the device Hamiltonian with first nearest neighbor interaction \cite{SC1,SC2,LLG1,NLLG,EEST}. The device consists of $N$ lattice or discrete points along the transport direction ($\hat{z}$). This includes the channel/insulator region, represented by (N-4) points, along with one layer from each of the FM regions (2 points), and one layer at the interface between the ferromagnet and the insulator for both FM layers (2 points)\cite{QTransport}. The first nearest neighbor interaction allows the inclusion of a minimum of one layer from each FM region (2 points) in the device Hamiltonian\cite{QTransport}. However, in the case of higher-order nearest-neighbor interaction, an additional number of layers of ferromagnets are required. We describe the FM contacts using the Stoner model of ferromagnetism with FM exchange energy ($\delta$), effective mass ($m_{FM}^*$), and Fermi energy ($E_f$)\cite{SC1,SC2,LLG1,Qt_um_2022}, and insulating barrier (MgO) having a barrier ($U_B$) above the Fermi-energy and effective mass ($m_{c}^*$)\cite{STT,NS_2,Qt_um_2022}.\\
\indent We have adopted the band description of the FM-MgO-FM system as presented in Ref. \cite{NS_2,Qt_um_2022}. To facilitate numerical implementation, we rewrite Eq.~\ref{eq:SecQ} using the notation adopted in Ref.~\cite{NEGF}. Along with that,  in this notation, the $H_0$ part of the Hamiltonian is an $N \times N$ block matrix, expressed as 
\begin{equation}
\label{eq:Hamil}
    \left[H_0\right]=\left(\begin{array}{cccccccc}
H_B^{F M} & \tau_B^{F M} & 0 & . & . & . & . & . \\
\tau_B^{\dagger F M} & H_I^{B \rightarrow C} & \tau_C & 0 & . & . & . & . \\
0 & \tau_C^{\dagger} & H^C & \tau_C & 0 & . & . & . \\
. & 0 & \tau_C^{\dagger} & H^C & \tau_C & 0 & . & . \\
. & . & 0 & . & . & . & 0 & . \\
. & . & . & 0 & . & . & . & 0 \\
. & . & . & . & 0 & \tau_C^{\dagger} & H_I^{C \rightarrow T} & \tau_T^{F M} \\
& . & . & . & . & 0 & \tau_T^{\dagger F M} & H_T^{F M}
\end{array}\right)
\end{equation}
where, each block has dimensions $P \times P$. Here, $P = 2LM$, where, $L$ and $M$ denote the number of points along the $\hat{x}$ and $\hat{y}$ directions, respectively. The factor of 2 accounts for the spin degrees of freedom. The matrix $H^{\text{FM}}_{B/T}$ represents the onsite energy matrix for the FM bottom/top contact, with $\hat{z}$ as the transport direction and $\hat{x}$ \& $\hat{y}$ as the transverse direction. It can be written as
\begin{subequations}
\label{eq:Hamill}
    \begin{equation}
         H_{B/T}^{F M}=(H_t^{F M} +2\operatorname{t_m}\textbf{I}_x \otimes \textbf{I}_y)\otimes \textbf{I}_2 +\Delta_{B/T}
    \end{equation}
    \begin{equation}
        H_t^{F M}=H^{FM}_x \otimes \textbf{I}_y +\textbf{I}_x \otimes{H}^{FM}_y 
    \end{equation}
    \end{subequations}
\noindent here, $\textbf{I}_x$, $\textbf{I}_y$, and $\textbf{I}_2$ are identity matrices of order $L \times L$, $M \times M$, and $2 \times 2$, respectively. The matrix $H_{x/y}$ represent the discretized tight-binding Hamiltonians of order $L/M \times L/M$ along the $\hat{x}$/$\hat{y}$ direction and given by
\begin{equation}
    \label{eq:HamilF_x}
    [H^{FM}_{x/y}]_{i,j}= \begin{cases}2t_m, & i=j, \\ -t_m, & i=j\pm1, \\ 0, & \text { else. }\end{cases}
\end{equation}
\noindent where, $t_m=\frac{\hbar^2}{2 m_{\text{FM}}^* a^2}$, $\hbar$ is the reduced Planck constant, $m_{\text{FM}}^*$ is the effective mass of the FM material, and $a$ is the lattice spacing or discretization step. The exchange energy matrix, $\Delta_{B/T}$, of the bottom/top FM layer with magnetization texture can be expressed as
\begin{equation}
  \Delta_{B/T} = \delta \dfrac{\textbf{I}_x \otimes \textbf{I}_y\otimes \textbf{I}_2 -M_{B/T}}{2} 
\end{equation}
where, $\delta$ is the ferromagnetic exchange energy. The $M_{B/T}$ is a block diagonal matrix of order $P \times P$, representing the spatial variation of the magnetization of the bottom/top contact. The diagonal entries of this matrix are specified by
\begin{equation}
    [M_{B/T}]^{ij}_{ij} = \hat{m}(x_i,y_j, z_{B/T}).\vec{\sigma}
\end{equation} 
where, $[M_{B/T}]^{ij}_{ij}$ is a $2 \times 2$ matrix, $\vec{\sigma}$ are Pauli matrices, and $\hat{m}(x_i, y_j, z_{B/T})$ is the spatially varying unit vector of magnetization of the bottom/top FM layer. The $\tau^{FM}_{B/T}$ is the coupling matrix of the bottom/top FM layer, written as
\begin{equation}
    \label{eq:tau_B_T_FM}
  \tau^{FM}_{B/T} = -t_m(\textbf{I}_x \otimes \textbf{I}_y \otimes  \textbf{I}_2).
\end{equation}
\indent Similarly, $H^C$ is the insulator onsite energy matrix, given by
\begin{subequations}
    \label{eq:Hamil_c}
    \begin{equation}
        H^{C}=(H_t^{C} +2\operatorname{t_c}\textbf{I}_x \otimes\textbf{I}_y)\otimes \textbf{I}_2
    \end{equation}
    \begin{equation}
        H_t^{C}=H^{C}_x \otimes \textbf{I}_y +\textbf{I}_x \otimes{H}^{C}_y 
    \end{equation}
\end{subequations}
where, $t_c=\frac{\hbar^2}{2 m_{c}^* a^2}$ represents the site-to-site hopping energy of the insulator with an effective mass $m^*_c$. The coupling matrix of the insulator region is written as 
\begin{equation}
    \label{eq:tau_C}
            \tau_{C} = -t_c(\textbf{I}_x \otimes \textbf{I}_y \otimes \textbf{I}_2). 
\end{equation}
\indent The $H^{B \rightarrow C}_I$ and $H^{C \rightarrow T}_I$ are the on-site energy matrices of the interface between the insulator and the FM layer, expressed as 
\begin{subequations}
    \label{eq:Hamil_I}
    \begin{equation}
         H_I^{B \rightarrow C/C \rightarrow T}=\left[H_t^{I} +(t_c+t_m)\textbf{I}_x \otimes \textbf{I}_y \right] \otimes \textbf{I}_2
    \end{equation}
\begin{equation}
        H_t^{I}=H^{I}_x \otimes \textbf{I}_y +\textbf{I}_x \otimes{H}^{I}_y 
\end{equation}
\end{subequations}
$H^{I}_{x/y}$ is the tight-binding Hamiltonian of order $L/M \times L/M$ along the $\hat{x}$/$\hat{y}$ direction at the interface, given by
\begin{equation}
    [H^{I}_{x/y}]_{i,j}= \begin{cases}t_m+t_c, & i=j, \\ -(t_m+t_c)/2, & i=j\pm1, \\ 0, & \text { else. }\end{cases}
\end{equation}
\indent Finally, the potential energy matrix, which includes both the applied bias and barrier potential (\(v_i\)), is expressed as
\begin{equation}
    [V]_{i,j}= \begin{cases}v_i \cdot \textbf{I}_x \otimes \textbf{I}_y \otimes \textbf{I}_2 , & i=j, \\ 0, & \text { else. }\end{cases}
\end{equation}
\indent To calculate the current and relevant observables, we use the NEGF method, where the retarded Green's function at energy (E) is given by
\begin{align} 
    G(E) = [(E+i\eta)\textbf{I} - H - \Sigma^r(E)]^{-1}
\label{eq:Gret}
\end{align}
where, $\eta$ is positive infinitesimal, $\Sigma^r$ is the self-energy matrix\cite{NEGF} of order $N \times N$ and can be articulated as
\begin{equation}
    \Sigma^r_{i,j}= \begin{cases}\Sigma_B, & i=j=1, \\ \Sigma_T, & i=j=N, \\ \textbf{0}, & \text { else }\end{cases}
\end{equation}
here, $\Sigma_{B/T}$ is self energy matrix of bottom/top contact with order $P\times P$. The current can be calculated using the transmission matrix($\hat{T}(E)$) as
\begin{equation}
     I = \frac{q^2}{\hbar} \int dE \, \operatorname{Re}\left[\operatorname{Tr}\left({\hat{T}}(E)\right)(f_T(E)-f_B(E))\right]
     \label{eq:Itr}
\end{equation}
where, $q$ is the charge of the electron, \( f_B(E) \) and \( f_T(E) \) are the Fermi–Dirac distributions of the bottom and top FM contacts, respectively. The transmission operator is given by  
\begin{equation}
    \hat{T}= \tilde{\Gamma}_T\,G\,\tilde{\Gamma}_B \, G^\dagger
\end{equation}
here, $\tilde{\Gamma}_{B/T}$ is the broadening matrix of bottom/top contact. For our device, the broadening matrix of top and bottom contact is given by $\tilde{\Gamma}_T= (\Gamma_T \otimes E_{NN})$  and $\tilde{\Gamma}_B=(\Gamma_B \otimes E_{11})$ respectively. Here, $E_{11}$ \& $E_{NN}$ are the matrix unit matrices of the order $N\times N$ and $\Gamma_{T/B}$ is same as define as
\begin{equation}
    \label{eq:gamma_BT}
    \Gamma_{B/T} =i(\Sigma_{B/T} - \Sigma^\dagger_{B/T}).
\end{equation}

Since trace operation is involved in the current calculation, the transmission operator can be written as
\begin{equation} \hat{T}= \left(\Gamma_T\right)\left(G_{N,1}\right)\left(\Gamma_B\right)\left(G_{N,1}\right)^{\dagger}
\label{eq:Top}
\end{equation}
here, \( G_{N,1} \) denotes the element (a sub-matrix of order \( P \times P \)) at the \( N \)th row (last row) and 1st column of the retarded Green's function matrix (Eq.~\ref{eq:Gret}). But there is an alternative expression for current, expressed in terms of the current operator, which is more relevant in the case of spin current. The current operator, which encompasses both spin and charge currents, is expressed as
\begin{equation}
\hat{\Pi}(E) =\frac{i}{\hbar} \left( H_{N-1,N} G^n_{N,N-1} - \left( H_{N,N-1} G^n_{N-1,N} \right)^\dagger \right)
\label{eq:Iop}
\end{equation}
here, $H$ is Hamiltonian of the device and $G^n$ is less than green function define as :
\begin{equation}
    G^n = i G \, \tilde{\Gamma}_L \,G^{\dagger}\,f_L + i G \, \tilde{\Gamma}_R \, G^\dagger \,f_R
\end{equation}

The charge current and the spin current along any direction \( \vec{s} \) can be determined using the following expression:
\begin{eqnarray}
I &=& \frac{q^2}{\hbar} \int dE \, \operatorname{Re}\left[\operatorname{Tr}_2\left(\operatorname{Tr}_1\left(\hat{\Pi}(E)\right) \cdot \textbf{I}_2 \right)\right] \label{eq:Ic} \\ 
I_{\vec{s}} &=& \frac{q^2}{\hbar} \int dE \, \operatorname{Re}\left[\operatorname{Tr}_2\left(\operatorname{Tr}_1\left(\hat{\Pi}(E)\right) \cdot \hat{S} \right)\right] \label{eq:Is}
\end{eqnarray}
here, \( \hat{S} = \vec{\sigma} \cdot \vec{s} \) and the first trace, \( \operatorname{Tr}_1 \), is taken over non-spin degrees of freedom, while the second trace, \( \operatorname{Tr}_2 \), is taken over spin degrees of freedom. For 3D structures, particularly where the contact properties vary along the transverse direction—such as MTJs with magnetic textured contacts—calculating the retarded Green's function, and thus the current, becomes computationally challenging. This is primarily due to the large size of the Hamiltonian matrix (\(N \cdot P \times N \cdot P\)), making it potentially infeasible\cite{Number}. In the following sections, we lay the groundwork to address the same while preserving the underlying physics of the device. The observation and subsequent simplifications presented in the following section are quite general in nature and applicable to a wide variety of devices. Hence, the method we develop can be seamlessly extended to a broad range of problems in quantum transport \cite{Number}, including those involving NEGF along with DFT calculations and time-dependent quantum transport calculations \cite{DFT1,ACNEGF1,ACNEGF2}.

\subsection{\label{sec:Decoupled}Uniform magnetization and decoupled spatio-eigen approach of NEGF}
\begin{figure}[htbp]
\centering
\includegraphics[width=\columnwidth]{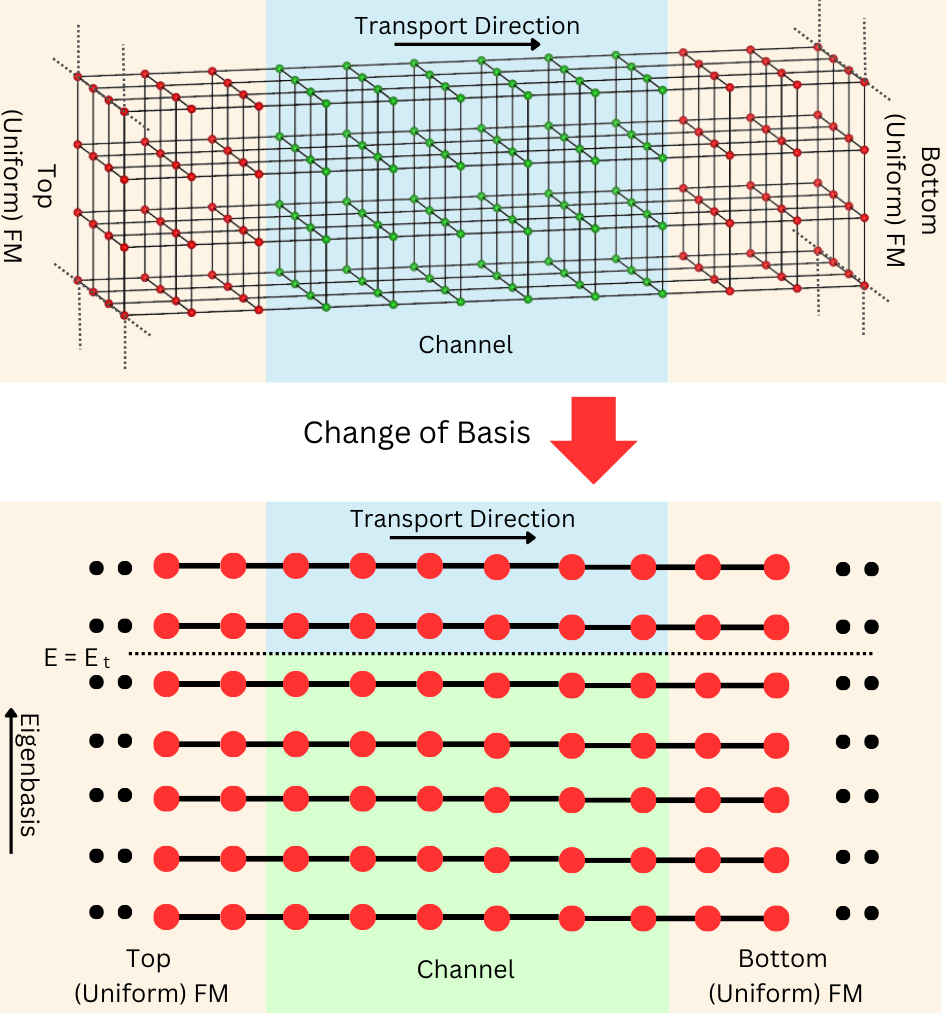}
\caption{Transformation of u-MTJ from tight-binding basis to spatio-eigen basis.}
\label{fig:Unifrom}
\end{figure}

We first outline the decoupled spatio-eigen approach of NEGF to address the formidable computational demand of solving large 3D systems such as u-MTJs and establish the concept of the limited bandwidth of incoming electron excitation. If both the FMs have uniform magnetization, then their exchange energy matrix can be written as
\begin{equation}
    \Delta_{B/T} =\textbf{I}_x \otimes\ \textbf{I}_y \otimes \left[ 
    \delta \left(\dfrac{\textbf{I}_2-\hat{m}_{B/T}.\vec{\sigma}}{2}\right) \right] 
    \label{eq:displit}
\end{equation}
here, $\hat{m}_{B/T}$ is the direction of magnetization of bottom/top contact, and $\delta$ is exchange splitting energy. For the case of uniform magnetization, the transverse Hamiltonians $H^{FM}_{B}$ and $H^{FM}_{T}$ (see Eq.~\ref{eq:Hamill}) can be block-diagonalized up to spin-dependent components. The Hamiltonian terms that are independent of spin-splitting, namely $H_t^{FM}$, $H^{C}_t$, and $H^{I}_t$ in Eqs.~\ref{eq:Hamill}, \ref{eq:Hamil_c}, and \ref{eq:Hamil_I}, can be simultaneously diagonalized, with their eigenvalues represented by $\epsilon^{FM}_i$, $\epsilon^{C}_i$, and $\epsilon^{I}_i$, respectively.\\
\indent The off-diagonal elements in the tight-binding Hamiltonian (Eq.~\ref{eq:HamilF_x}) along the transverse direction represent interactions between adjacent lattice points in that direction. By simultaneously diagonalizing the transverse Hamiltonian matrices—specifically \( H_t^{FM} \), \( H_t^{C} \), and \( H_t^{I} \)—the device (u-MTJ) can be effectively partitioned into multiple decoupled, spin-dependent 1D channels, each corresponding to distinct eigenvalues in the transverse direction (see Fig.~\ref{fig:Unifrom}). This approach enables the NEGF equations to be formulated in a decoupled spatio-eigen basis, allowing each 1D channel to be treated independently, as illustrated schematically in Fig.~\ref{fig:Unifrom}.\\
\noindent In this context, the Hamiltonian for each of these transverse eigen ($i$) channels is represented by
\begin{equation}
\left[H^i_0\right]=\left(\begin{array}{cccccccc}
\alpha_B^{F M} & \beta_B^{F M} & 0 & . & . & . & . & . \\
\beta_B^{\dagger F M} & \alpha_I^{B \rightarrow C} & \beta_C & 0 & . & . & . & . \\
0 & \beta_C^{\dagger} & \alpha_C & \beta_C & 0 & . & . & . \\
. & 0 & \beta_C^{\dagger} & \alpha_C & \beta_C & 0 & . & . \\
. & . & 0 & . & . & . & 0 & . \\
. & . & . & 0 & . & . & . & 0 \\
. & . & . & . & 0 & \beta_C^{\dagger} & \alpha_I^{C \rightarrow T} & \beta_T^{F M} \\
& . & . & . & . & 0 & \beta_T^{\dagger F M} & \alpha_T^{F M}
\end{array}\right)
\end{equation}
where, $\alpha_B^{F M}$, $\alpha_I^{B \rightarrow C}$, $\alpha^C$, $\alpha_I^{C \rightarrow T}$, and $\alpha_T^{F M}$ are spin-dependent on-site energy matrices of the bottom FM, interface between the bottom FM and the insulator region, the insulator, the interface between the insulator region and the top FM, and the top FM contact, respectively, given by
\begin{subequations}
    \begin{equation}
        \alpha_{B/T}^{F M}=\left(2 t_m+\epsilon_i^{F M}\right) \textbf{I}_2+\Delta^{2\times2}_{B/T}
    \end{equation}
    \begin{equation}
        \alpha_I^{B \rightarrow C/C \rightarrow T}=\left(t_c+t_m+\epsilon_i^{I}\right)\textbf{I}_2
    \end{equation}
    \begin{equation}
        \alpha^{C}=\left(2 t_c+\epsilon_i^{C}\right)\textbf{I}_2
    \end{equation}
 \end{subequations}
\noindent where, the spin-splitting energy matrix for these transverse eigen ($i$) channels is given by $\Delta^{2\times2}_{B/T}=\delta \frac{\textbf{I}_{2}-\hat{m}_{B/T}\cdot\vec{\sigma}}{2}$, as shown in Eq.~\ref{eq:displit}. The matrices $\beta^{FM}_{B/T}$ and $\beta^{C}$ are the coupling matrices of the bottom/top FM contact and insulator, respectively, given by
\begin{subequations}
    \begin{equation}
        \beta^{FM}_{B/T} = -t_m\textbf{I}_2
    \end{equation}
\begin{equation}
  \beta^{C} = -t_c\textbf{I}_2.
\end{equation}
\end{subequations}

Finally, we have a potential energy matrix, including basing potential and barrier potential 
\begin{equation}
    [V]_{i,j}= \begin{cases}v_i*\textbf{I}_2, & i=j, \\ 0, & \text { else. }\end{cases}
\end{equation}

Now, this recasting of the Hamiltonian ($H_0$) allows one to solve NEGF for each transverse eigen ($i$) 1D-channel independently. The non-equilibrium retarded Green's function of each channel ($i$) can be written as
\begin{equation}
    [G^i(E)]=[(E+i\eta) \textbf{I}-(H^i_0+V)-\Sigma^i]^{-1},
\end{equation}
where, the self energy matrix $\Sigma^i$ is given by
\begin{equation}
    \Sigma_{m n}^i= \begin{cases}\Sigma_B^i, & m=n=1, \\ \Sigma_T^i, & m=n=N, \\ 0_{2,2}, & \text { else. }\end{cases}
\end{equation}
In general, self energy can be evaluated using the surface Green's function, but in the case of a 1D chain with the reflection-less contact with an open-boundary condition, each term of the self-energy in this can be written as \cite{NEGF}
\begin{equation}
        \Sigma_{B/T}^i=-t_m R_{B/T}\left[\begin{array}{cc}\exp \left(i k_{B/T}^{\uparrow,i}a \right)& 0 \\0 & \exp \left(i k_{B/T}^{\downarrow,i} a\right)\end{array}\right] R_{B/T}^{\dagger}
\end{equation}
where, $R_{B/T}$ is a spinor rotation matrix that rotates self-energy matrices from $\hat{z}$ to along the direction of magnetization of the bottom/top contact. The $k_{B/T}^{\uparrow/\downarrow,i}$ is related to the spin-dependent $E-k$ relation inside the FM for each transverse eigen($i$) channel as
\begin{subequations}
\label{eq:EK}
    \begin{equation}
        E =\epsilon^{FM}_i+v_{T} +2 t_{m}\left(1-\cos k_{B/T}^{\uparrow,i} a\right),
    \end{equation}
    \begin{equation}
        E =\epsilon^{FM}_i+\delta+ v_{B}+ 2 t_{m}\left(1-\cos k_{B/T}^{\downarrow,i} a\right)
    \end{equation}
\end{subequations}
where, $v_{B/T}$ is the potential at the bottom/top contact. The broadening matrix is given by 
\begin{equation}
        \Gamma_{B/T}^i=-2t_m R_{B/T}\left[\begin{array}{cc}\cos \left( k_{B/T}^{\uparrow,i}a \right)& 0 \\0 & \cos \left( k_{B/T}^{\downarrow,i} a\right)\end{array}\right] R_{B/T}^{\dagger}
    \label{eq:Gai}
\end{equation}
\indent The charge and spin current (Eq. \ref{eq:Iop}) can be calculated by adding the current of each individual channel as
\begin{equation}
\label{eq:Ci}
I = \frac{q^2}{\hbar} \int dE \,\sum_i \operatorname{Re}\left[\operatorname{Tr}_2\left(\operatorname{Tr}_1\left(\hat{\Pi^i}(E)\right)\right)\right]
\end{equation}
\indent The major computational advantage of the spatio-eigen approach arises from the fact that it is not necessary to solve the NEGF for all eigen ($i$) 1D-channels to capture the current. It can be inferred from Eq.~\ref{eq:EK} that for a specific energy ($E$), if:
\begin{equation}
\label{eq:condition}
    \lvert 1-\frac{E- \delta - \epsilon^{FM}_i - u_{B/T}}{2t_m}\rvert > 1
\end{equation}

then, $k_{B/T}^{\uparrow/\downarrow,i}$ becomes purely imaginary, reducing $[\Gamma^i_T]$ and $[\Gamma^i_B]$ to zero (Eq.~\ref{eq:Gai}). Consequently, the current flow from the contacts to the insulator (Eq.~\ref{eq:Ci}) vanishes. As a result, the higher transverse eigen ($i$) channel does not contribute to the current flow, thereby restricting any NEGF calculation to a limiting range of transverse eigenvalues ($\epsilon_i^{FM} < E_t$; see Fig.~\ref{fig:Unifrom}). It is worth noting that we have used the current operator as an archetype for our discussion on the spatio-eigen NEGF framework; however, the same applies to other operators, such as transmission, spectral density, and correlation operators.\\
\indent There is a crucial inference to be made that will aid in solving the NEGF for the Sk-MTJ and other devices. The vanishing current at higher transverse energy levels can be explained by the fact that these eigenvalues associated with specific transverse eigenbasis represent the electron's energy in the transverse direction. As a result, when the transverse energy $\epsilon^{FM}_i$ exceeds $E$ (see Eq.~\ref{eq:EK}), electrons originating from the contact do not carry crystal momentum along the transport direction, leading to a diminished contribution to the current in the higher transverse eigen-channels after a certain threshold. \\
\indent The diminishing nature of current in the higher transverse eigen-energy is independent of the specific dispersion relations or self-energy terms employed in the model. Even when considering more complex Hamiltonians and self-energies, the current and the broadening matrices tend to decrease as the transverse component of energy increases. This understanding drives the development of a more efficient approach for NEGF-based quantum transport, focusing exclusively on contributions from the lower transverse eigen ($i$) channels to enhance computational efficiency.\\
\indent The proposed spatio-eigen method for NEGF may offer a potential solution to alleviate the high computational demand associated with coupled DFT and quantum transport calculations. Traditional DFT + NEGF approaches \cite{DFT2,DFT3} or quantum transport methods rely on the Bloch expansion of the device Hamiltonian and self-energies over the transverse momentum to accurately capture transport features. However, the significant computational cost arises from the need for a dense \textit{k}-grid (2-dimensional for 3D structures) and a comprehensive sampling of the entire transverse Brillouin zone to account for transport features such as transmission, current, and spectral density. In contrast, the spatio-eigen framework for NEGF, when combined with DFT, can substantially reduce computational complexity by transforming transport features into vanishing functions at higher transverse eigen-energies.
\subsection{\label{sec:Coupled}Micromagnetic texture and coupled spatio-eigen approach of NEGF}
The decoupled spatio-eigen approach of NEGF encounters a major limitation when applied to MTJs with ferromagnetic contacts featuring micromagnetic textures, such as magnetic skyrmions. This challenge arises because the transverse Hamiltonians of the contacts, labeled \( H^{FM}_B \) and \( H^{FM}_T \), cannot be simultaneously block-diagonalized into reduced-order matrices (as shown in Eq.~\ref{eq:Hamill}). Consequently, this non-commutative nature undermines the decoupled spatio-eigen approach of NEGF. Nonetheless, the core concept outlined earlier—namely, the diminishing impact of higher transverse eigen channels on current—remains advantageous for efficiently calculating NEGF in MTJs with non-uniform magnetization textures. It's worth noting that previous efforts, such as by P. Flauger et al. \cite{Qt_um_2022}, aimed to bridge quantum transport and micromagnetism, have overlooked the simultaneous diagonalization of the transverse Hamiltonians. This omission limits their approach's applicability primarily to larger micromagnetic structures. In contrast, the method proposed here accommodates any Hamiltonian, irrespective of the transverse direction symmetry. \\
\indent In this section, our aim is to simplify the NEGF calculation for an MTJ with any micromagnetic texture by focusing solely on the conducting transverse eigenchannels, thereby reducing the matrix size while preserving the underlying physics. For that, we transform the transverse Hamiltonians of the device—encompassing both contacts and the insulator—along with the coupling matrices (see Eqs.~\ref{eq:tau_B_T_FM} \& \ref{eq:tau_C}) into the eigenbasis of the top contact's transverse Hamiltonian \( H_T^{FM} \). Although the transverse Hamiltonians for the bottom (\( H_B^{FM} \)) and top (\( H_T^{FM} \)) contacts cannot be diagonalized simultaneously, it is possible to diagonalize the transverse Hamiltonians of the insulator (\( H^{C} \)) and the top contact (\( H_T^{FM} \)) together. Let the simultaneous transverse eigenbasis of \( H_T^{FM} \) and \( H^{C} \) be represented by \( |\epsilon_i\rangle \). In this transverse eigenbasis, \( H_T^{FM} \) and \( H^{C} \) transform into their respective diagonal eigen matrices, \( \epsilon^{FM}_T \) and \( \epsilon^{C} \), respectively, while the coupling matrices (see Eq. \ref{eq:tau_B_T_FM},\ref{eq:tau_C}) remain unchanged. However, because \( H_B^{FM} \) and \( H_T^{FM} \) do not commute, transforming \( H_B^{FM} \) gives a non-diagonal matrix in the transverse eigenbasis. Let's say $\epsilon^{FM}_B$ is the eigenvalue matrix of bottom contact, then the transverse Hamiltonian of bottom contact in simultaneous transverse eigenbasis of top contact and insulator is given by 
\begin{equation}
    h^{FM}_B = \textbf{U}^\dagger\epsilon^{FM}_B\textbf{U}
\end{equation}
\begin{equation}
\label{eq:Uij}
    \textbf{U}_{i,j} = \langle\epsilon'_i|\epsilon_j\rangle
\end{equation}
where, $|\epsilon'_i\rangle$ and $|\epsilon_i\rangle$ are the eigenbasis of the bottom contact and the simultaneous transverse eigenbasis, respectively.\\
\begin{figure}[htbp]
\centering
\includegraphics[width=\columnwidth]{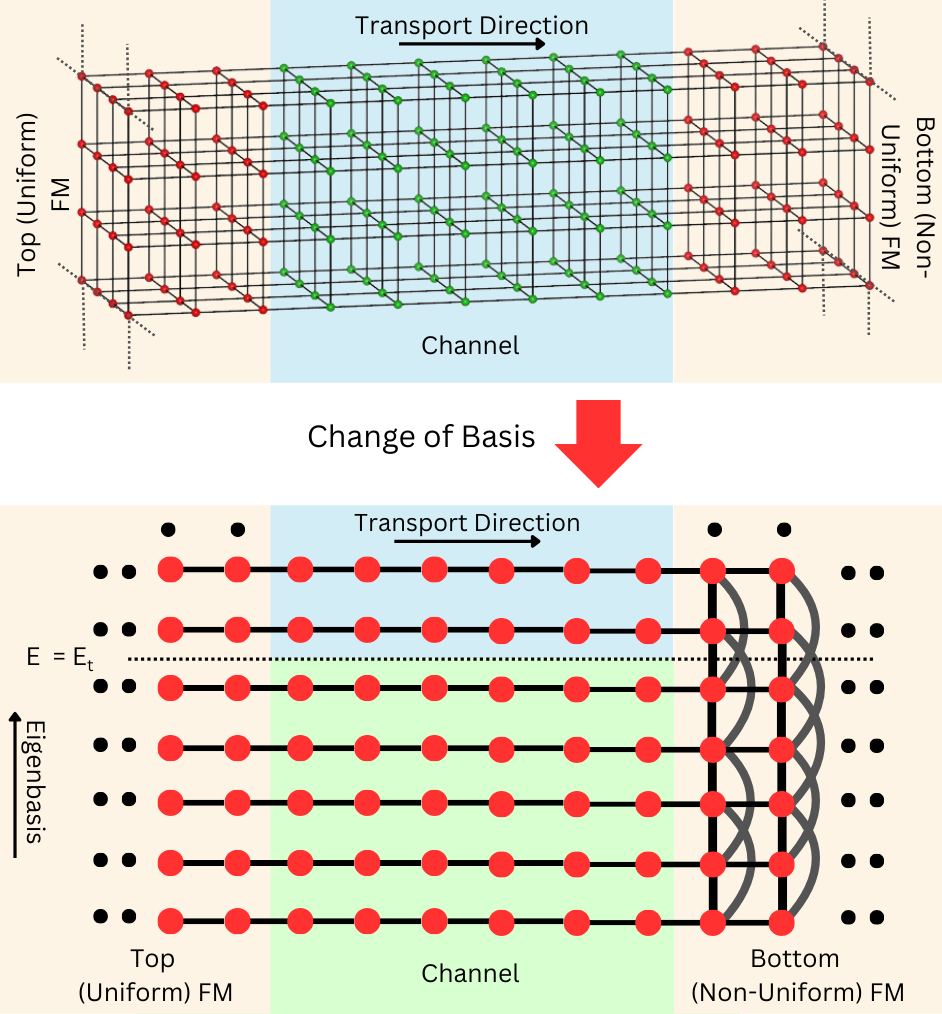}
\caption{Transformation of Sk-MTJ from tight-binding space basis to spatio-eigen basis.}
\label{fig:NonUnifrom}
\end{figure}
\indent Since the coupling matrices (see Eq. \ref{eq:tau_B_T_FM},\ref{eq:tau_C}) remain intact in the transverse eigenbasis, the device in the spatio-eigen approach of NEGF can be envisioned as shown in the Fig.~\ref{fig:NonUnifrom}. The self-energy matrix for the top contact in the transverse eigenbasis takes diagonal form, articulated as \cite{NEGF2}
\begin{equation}
\label{eq:sigL}
    \Sigma_{T(i,j)}= \begin{cases}-t_m \exp\left(i k^i_Ta \right), & i=j, \\ 0, & \text { else. }\end{cases}
\end{equation}
\begin{equation}
    E =\epsilon^{FM}_{T,i}+v_{T} +2 t_{m}\left(1-\cos k^i_{T}a\right).
\end{equation}
here, \( \epsilon^{FM}_{T,i} \) denotes the transverse eigenvalue of the top contact. The broadening matrix, represented as \( [\Gamma_T]_{i,i} = [\Sigma_{T}]_{i,i} - [\Sigma^{\dagger}_{T}]_{i,i} \), also assumes a diagonal form. In a decoupled scenario, it is observed that the term \( e^{i k_T a} \) becomes purely real beyond a certain transverse eigenvalue. The observation remains intact in the coupled case also, and the corresponding entries for these transverse energies in the broadening matrix reduce to zero after a certain transverse energy threshold ($E_t$; see Fig.~\ref{fig:NonUnifrom}). This results in two distinct categories of eigenvalues: a `relevant' set that contributes to transmission, and an `irrelevant' set that does not play a role in conduction. As a result, the eigenvalue matrix of the top FM's transverse Hamiltonian ($H^{FM}_T$) can be structured into block matrices of size \( 2 \times 2 \) as
\begin{equation}
\label{eq:Split}
\begin{array}{l}
\epsilon^{FM}_T =\left(\begin{array}{ll}
\epsilon^{T}_{rev} & 0 \\
0 & \epsilon^{T}_{irr}
\end{array}\right)_{\lambda}
\end{array}
\end{equation}
here, $\epsilon^{T}_{rev}$ encompasses all the transverse eigenvalues of the top contact below a threshold transverse energy (will refer as $E_t$), beyond which both $e^{i k_T a}$ and $e^{i k_B a}$ become purely real. The remaining transverse eigenvalues are included in $\epsilon^{T}_{irr}$. We will use the $\lambda$ subscript for matrices that are partitioned in this manner. A similar partition is applied to the transverse eigenvalues of the bottom contact. With this separation, the broadening matrix of the top contact is simplified to:
\begin{equation}
\label{eq:gamL}
\begin{array}{l}
\Gamma_T =\left(\begin{array}{ll}
\Gamma^T_{rel} & 0 \\
0 & 0
\end{array}\right)_{\lambda.}
\end{array}
\end{equation}
\indent The self-energy matrix for the bottom contact can be derived by first writing the self-energy in the transverse eigenbasis of the bottom contact (\( \Sigma_B^{'} \)) and then transforming it to the simultaneous transverse eigenbasis of the top contact and insulator. This transformation is achieved using a unitary transformation, described by the Eq.~\ref{eq:Uij}, as
\begin{equation}
    \Sigma_B = \textbf{U}^\dagger\Sigma_B^{'}\textbf{U}
\end{equation}
where, $\Sigma^{'}_B$ is self energy of the bottom contact in $|\epsilon'_i\rangle$ eigen basis, given by
\begin{equation}
\label{eq:sigB'}
    \Sigma^{'}_{B(i,j)}= \begin{cases}-t_m\exp\left(i k^i_B a \right), & i=j, \\ 0, & \text { else. }\end{cases}
\end{equation}
\begin{equation}
    E =\epsilon^{FM}_{B,i}+v_{B} +2 t_{m}\left(1-\cos k^i_{B}a\right).
\end{equation}
where, $\epsilon^{FM}_{B,i}$ is transverse eigen value of the bottom contact. Employing the same rationale as presented in the equation in Eq. \ref{eq:gamL}, the broadening matrix of the bottom contact in the transverse eigenbasis can be written as
\begin{equation}
    \begin{array}{l}
\Gamma_B= \textbf{U}^\dagger\left(\begin{array}{ll}
\Gamma^{'B}_{rev} & 0 \\
0 & 0
\end{array}\right)_{\lambda}\textbf{U}.
\end{array}
\end{equation}
\indent On the similar lines, we partition the matrix $\textbf{U}$ into block matrices akin to $\Gamma$ and $\Sigma$, the aforementioned equation can be expressed as
\begin{equation}
    \begin{array}{l}
\Gamma_B=\left(\begin{array}{cc}
U_{11}^{\dagger} & U_{21}^{\dagger} \\
U_{12}^{\dagger} & U_{22}^{\dagger}
\end{array}\right)_{\lambda}\left(\begin{array}{ll}
\Gamma^{'B}_{rel} & 0 \\
0 & 0
\end{array}\right)_{\lambda}\left(\begin{array}{ll}
U_{11} & U_{12} \\
U_{21} & U_{22}
\end{array}\right)_{\lambda} \\
\end{array}
\end{equation}
which, simplifies to
\begin{equation}
\label{eq:g2}
\begin{array}{l}
\Gamma_B =\left(\begin{array}{ll}
U^{\dagger}_{11}\Gamma^{'B}_{rel}U_{11}  & U^{\dagger}_{11}\Gamma^{'B}_{rel}U_{12} \\
U^{\dagger}_{12}\Gamma^{'B}_{rel}U_{11} & U^{\dagger}_{12}\Gamma^{'B}_{rel}U_{12}
\end{array}\right)_{\lambda}
=\left(\begin{array}{ll}
\Gamma_{11} & \Gamma_{12} \\
\Gamma_{21} & \Gamma_{22}
\end{array}\right)_{\lambda}
\end{array}
\end{equation}
\indent Similarly, the matrix $G_{N,1}$ can be represented in a $2 \times 2$ block matrix as
\begin{equation}
\label{eq:gn_combined}
\begin{array}{cc}
G_{N,1} = \left(\begin{array}{ll}
g^a_{11} & g^a_{12} \\
g^a_{21} & g^a_{22}
\end{array}\right)_{\lambda}
\end{array}
\end{equation}
\indent Since only trace of operators is involved in the current calculation (Eq.\ref{eq:Itr}, \ref{eq:Is}), hence, using Eq.~\ref{eq:gamL}, \ref{eq:g2} \& \ref{eq:gn_combined}, the relevant part of transmission operator in Eq.~\ref{eq:Top} can be written as
\begin{equation}
\begin{aligned}
\hat{T}_{\text{eff}}(E) &= \Gamma^{T}_{\text{rel}} \, g^a_{11}\, \Gamma_{11} \,g^{a,\dagger}_{11} \,+\quad \Gamma^{T}_{\text{rel}}\, g^a_{11}\, \Gamma_{21}\, g^{a,\dagger}_{21} \,+ \\
&\quad \Gamma^{T}_{\text{rel}} \,g^a_{21}\, \Gamma_{12}\, g^{a,\dagger}_{12} \,+\quad \Gamma^{T}_{\text{rel}}\, g^a_{21}\, \Gamma_{22}\, g^{a,\dagger}_{22}
\end{aligned}
\end{equation}

To summarize, by doing this, we effectively view the device as a series of 1D transverse eigen($i$) channels. However, these eigenchannels are now interconnected at the bottom contact's end as \textbf{$H^{FM}_{B}$} have non-diagonal terms as well (Fig. \ref{fig:NonUnifrom}). If there are additional non-uniformity, they may lead to further interconnections. If we examine \(\Gamma_B\) (Eq. \ref{eq:g2}), unlike \(\Gamma_T\), its diagonal block corresponding to irrelevant channels and its off-diagonal blocks are non-zero, indicating conduction in irrelevant channels and between channels as well. This conduction arises from the overlap of some eigenbasis of the irrelevant section with those of the relevant section, as represented by \(U_{12}\) and \(U_{21}\). Thus, even though only a limited number of channels contribute to conduction, these overlaps can lead to connections in higher channels. Our goal is to account for this effect while keeping the set of relevant channels small. To achieve this, we make the approximation that each eigen($i$) channel is meaningfully connected to a finite number of other eigen-channels. Another way to say this is for a given $i$, term $\textbf{U}_{i,j}$ (Eq. \ref{eq:Uij}) is meaningfully non-zero for a finite number of $j$s and vice versa. A detailed justification for this approximation is provided in appendix \ref{appA} and appendix \ref{appB}. The consequence of this assumption is that only a finite number of channels from the irrelevant section are connected to the relevant eigenvalues. Therefore, we can order the eigenvalues to group these non-zero entries of \( U_{12} \) together. If we further divide \( U_{12} \) and \( U_{21} \) into blocks of \( 2 \times 2 \) matrices, they are expected to have the following form:
\begin{subequations}
\label{eq:u12}
    \begin{equation}
    U_{12} = \left(\begin{array}{ll}
0 & 0 \\
u_1 & 0
\end{array}\right) \\
    \end{equation}

    \begin{equation}
    U_{21} = \left(\begin{array}{ll}
0 & u_2 \\
0 & 0
\end{array}\right) \\
    \end{equation}
\end{subequations}

The equation suggests that the primary overlap occurs between the higher energy eigenvectors of the relevant set and the lower energy eigenvectors of the irrelevant set. This overlap can result in minor conduction within the irrelevant set due to coupling with the conducting relevant set. To mitigate this effect, we have to ensure that these additional relevant eigenchannels, which overlap with the irrelevant set, do not contribute to conduction themselves. This can be achieved by increasing the threshold($E_t$) for relevant eigenvalues. As a result, specific diagonal entries of the relevant broadening matrix \(\Gamma^{B/T}_{\text{rev}}\) are set to zero. If \(u\) is an \(n \times n\) matrix, then we increase our threshold ($E_t$) to include \(n\) more eigenchannels. Since these eigenchannels do not conduct, this condition can be expressed as follows:
\begin{equation}
\label{eq:g4}
\begin{array}{l}

\Gamma^{'B/T}_{rev} =\left(\begin{array}{ll}
\Gamma & 0 \\
0 & 0
\end{array}\right).
\end{array}
\end{equation}
here, $\Gamma^{'B/T}_{rev}$ is also partitioned into block matrices, similar to $U_{12}$ and $U_{21}$ for the sake of clarity. Now, if we substitute the expressions from Eq. \ref{eq:g4} and Eq. \ref{eq:u12} into Eq. \ref{eq:g2}, it simplifies to:

\begin{subequations}
    \begin{equation}
\begin{array}{l}

\Gamma_{B} =\left(\begin{array}{ll}
\Gamma_{11} & 0 \\
0 & 0
\end{array}\right)_{\lambda}
\end{array}
\end{equation}

\begin{equation}
    \Gamma_{11} = \textbf{U}_{11}^{\dagger}\Gamma^B_{rel}\textbf{U}_{11.}
\end{equation}
\end{subequations}
\indent The contributing expression of transmission operator is reduced to:
\begin{align}
\hat{T}_{\text {eff}}
(E)&=\Gamma^{T}_{\text{rel}} \, g^a_{11}\, \Gamma_{11} \,g^{a,\dagger}_{11}
\end{align}
\indent Finally, we derive the expression for $g^a_{11}$ in terms of relevant components. We re-caste device Hamiltonian in spatio-eigen basis, i.e., transverse Hamiltonians and coupling matrices being in the simultaneous transverse eigenbasis of the top contact \& insulator. The inverse of the Green function can be represented as a $2 \times 2$ block matrix.
\begin{equation}
\label{eq:Grr}
\begin{array}{l}
G = [E\textbf{I} - H - \Sigma]^{-1}
=\left(\begin{array}{ll}
A & B \\
C & D
\end{array}\right)^{-1}
\end{array}
\end{equation}

\begin{subequations}
\label{eq:ABC}
    \begin{equation}
        A = U^\dagger(E\textbf{I} - \epsilon^{FM}_{B} - \Sigma^{\prime}_{B})U
    \end{equation}
    \begin{equation}
        B = [t_m\textbf{I}\ \textbf{0}\ \textbf{.}\ \textbf{.}\ \textbf{.}\ \textbf{0}]
    \end{equation}
    \begin{equation}
        C = B^{\dagger}
    \end{equation}
\end{subequations}
where, $D$ represents the rest of the matrix, which can be viewed as a block matrix of dimensions $N-1 \times N-1$, with each block having a dimension of $P \times P$, consistent with matrix A. Then, the matrix $G_{N,1}$ simplifies to (refer to the Appendix \ref{app:Mat}):
\begin{equation}
\label{eq:G1N}
G_{N,1}=  -t_m(D_{N-1,1}^{-1})(A-t_m^{2}(D_{1,1})^{-1})^{-1}
\end{equation}
\indent As all blocks of $D$ are diagonal matrices, both $(D^{-1})_{1,1}$ and $(D^{-1})_{N-1,1}$ are also diagonal matrices. Therefore, these matrices, along with matrices A, can be expressed in terms of relevant and irrelevant blocks as
\begin{subequations}
\label{eq:AD}
    \begin{equation}
\begin{array}{l}

A =\left(\begin{array}{ll}
a_{11} & a_{21} \\
a_{12} & a_{22}
\end{array}\right)_{\lambda}
\end{array}
\end{equation}

\begin{equation}
\begin{array}{l}

D_{1,1}^{-1} =\left(\begin{array}{ll}
d_{1} & 0 \\
0 & d_{2}
\end{array}\right)_{\lambda}
\end{array}
\end{equation}

\begin{equation}
\begin{array}{l}

D_{N-1,1}^{-1} =\left(\begin{array}{ll}
e_{1} & 0 \\
0 & e_{2}
\end{array}\right)_{\lambda}
\end{array}
\end{equation}
\end{subequations}
\indent Since, we are interested in $g^a_{11}$ (see Eq. \ref{eq:gn_combined}), we can expand the $G_{N,1}$ matrix into $2 \times 2$ blocks and replace A, $(D^{-1})_{11}$, and $(D^{-1})_{N-1,1}$ using Eq. \ref{eq:AD}. This allows us to obtain $g^a_{11}$ using the inverse of a $2 \times 2$ block matrix (refer to the Appendix \ref{app:Mat}).
\begin{equation}
g^a_{11}=  -t_m(e_1)(a_{11}-t_m^{2}d_1 -a_{12}(a_{22}-d_2)a_{21})^{-1}
\label{eq:gaf1}
\end{equation}
\indent Expanding the expressions of $a$'s using Eq. \ref{eq:ABC}.a, we obtain:
\begin{equation}
g^a_{11}=  -t_m(e_1)(U^{\dagger}_{11}(E\textbf{I}-\epsilon^{T}_B-\Gamma^{B}_{rel}) U_{11}-\tau_m^{2} d_1 + O(u^2))^{-1}
\label{eq:gaf2}
\end{equation}
\indent As shown in the analysis, the formula for \(g^a_{11}\) (and thus the current expression) is simplified to involve matrices associated only with relevant eigenvalues. However, a factor depending on higher-order terms related to matrix \(u\) persists. A similar exercise can be performed to show (Appendix \ref{CurrentOp}) that the current operator Eq. \ref{eq:Iop} also reduces to just the relevant terms, plus a term that is quadratic or of higher order in \(u\). As previously explained, matrix \(u\) characterizes the coupling between the higher relevant eigenchannels and the lower irrelevant channels. This factor decreases as the relevance threshold is raised, thereby reducing its impact. Consequently, the current expression can be constructed using only the relevant terms with an appropriate eigenvalue threshold. A demonstration of this claim using a 2D toy model is provided in Appendix~\ref{appA}, along with a justification of this approach for our device.\\
\indent The spatio-eigen framework of NEGF presented here eliminates the need to solve the full 3D NEGF while preserving the system's physics under non-equilibrium conditions. The substantial reduction in computational demand is achieved by fully constructing the current operator from a subset of the relevant transverse eigenchannels of the system. Notably, the spatio-eigen framework of NEGF is general and can handle \textit{ translationally variant} systems, provided the system's Hamiltonian can be derived from a tensor sum of Hamiltonians (see Eq.~\ref{eq:Hamill}).

\section{Modeling}
\label{sec:Model}
We considered the MTJ diameter of 20 nm with MgO thickness of 1 nm. We use a skyrmion with a diameter of 10 nm, stabilized on a material with saturation magnetization $M_s = 580 \, \mathrm{kA/m}$, exchange stiffness $A = 15 \, \mathrm{pJ/m}$, interfacial Dzyaloshinskii-Moriya interaction (DMI) constant $D = 3.5 \, \mathrm{mJ/m^2}$, and anisotropy constant $K_u = 1 \, \mathrm{MJ/m^3}$. The DMI from the bottom contact, which plays a key role in determining the skyrmion texture, is inherently accounted for in the micromagnetic calculations. Since the contact hosting the skyrmion is in local equilibrium with a chemical potential  $\mu_B$, we do not explicitly include spin-orbit coupling (SOC) in the Hamiltonian of the contact. Instead, SOC effects are incorporated within the magnetic interaction term. During the scaling of the MTJ, we have proportionally reduced the skyrmion size to understand the scaling effect, which may potentially be achieved via increased anisotropy or DMI \cite{SkyMD}. We have used $0.8 \, m_e$, $0.18 \, m_e$, $2.25 \, \mathrm{eV}$, $0.76 \, \mathrm{eV}$, and $2.15 \, \mathrm{eV}$ as the effective mass of electrons in MgO, FMs, the Fermi energy, MgO barrier height above Fermi level and the exchange splitting of FMs\cite{STT, NS_2,NLLG}, respectively. 
\begin{figure}
    \centering
    \includegraphics[width=\linewidth]{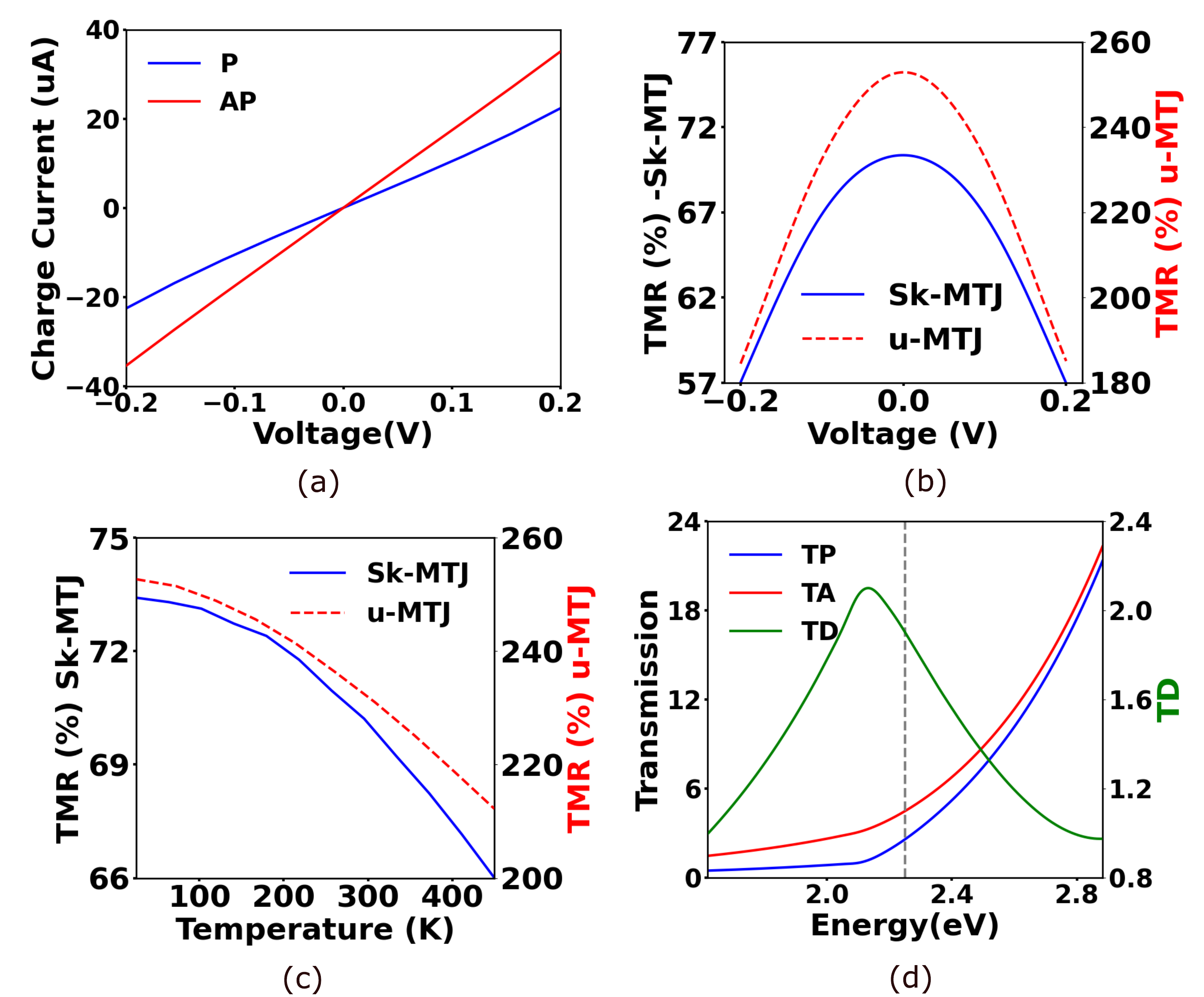}
    \caption{(a) Current-Voltage characteristics in the PC and the APC, (b) TMR variation with voltage, (c) TMR roll-off with temperature for both the Sk-MTJ and u-MTJ. (d) Transmission-spectra of the Sk-MTJ in the PC ($T_P$) and the APC ($T_{AP}$) along with the transmission difference ($T_D=T_P-T_{AP}$).}
  \label{fig:IV}
\end{figure}
\begin{figure}[t!]
    \centering
    \includegraphics[width=\linewidth]{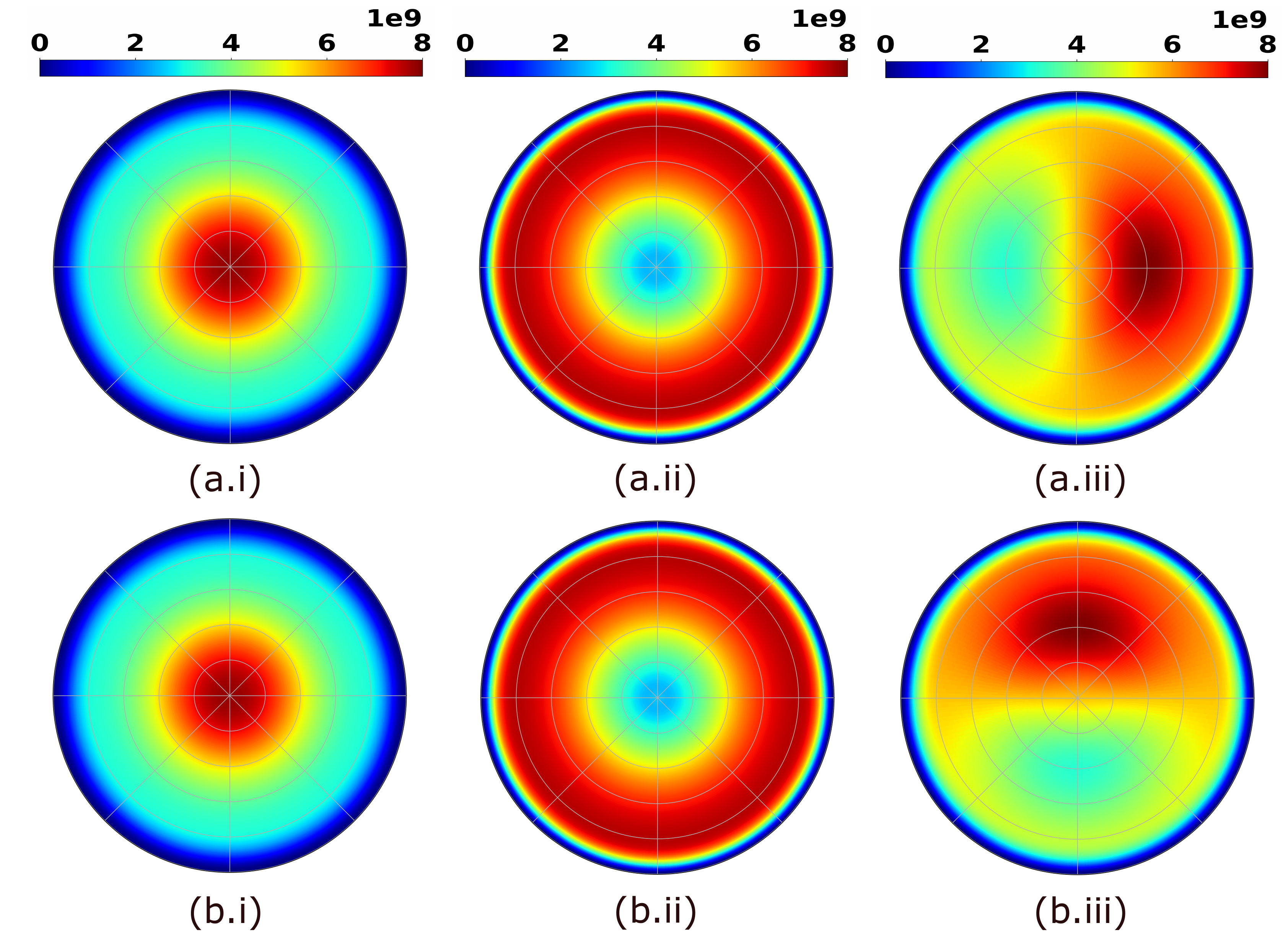}
    \caption{Charge current density($A/m^2$) profiles of the Sk-MTJ with u-FM oriented in (i) the PC, (ii) APC, and (iii) orthogonal (along $x$-axis) to the core of the skyrmion for (a) Néel and (b) Bloch-type skyrmion, at 10 mV.}
    \label{fig:Charge}
\end{figure}
\section{Results and discussion}
We begin our analysis by examining the current-voltage (I-V) characteristics of the Sk-MTJ as shown in Fig.~\ref{fig:IV}(a), in the parallel configuration (PC) and the anti-parallel configuration (APC). The PC/APC refers to the relative orientation of the top uniform (u)-FM with respect to the skyrmion core (Fig.\ref{fig:skyrmion}). Néel- and Bloch-type skyrmion in the Sk-MTJ exhibit identical I-V characteristics due to the equivalent projections of the skyrmion texture onto the u-FM in both the PC and the APC. The higher current observed in the APC compared to the PC is attributed to the more positive projection of the skyrmion texture onto the u-FM in the Sk-MTJ. Spin texture-dependent tunneling in the Sk-MTJ leads to higher resistance in the PC and lower resistance in the APC, quantified by the TMR as
\begin{equation}
T M R=\frac{R_{h}-R_{l}}{R_{l}} \times 100
\label{eqn:one} 
\end{equation}
here, $R_{h/l}$ denotes the high/low resistance in the PC/APC. The zero-voltage TMR of the Sk-MTJ is approximately 70\%, which is notably less than half the TMR of the u-MTJ (Fig.~\ref{fig:IV}(b)). Figure \ref{fig:IV}(c) illustrates a monotonic decrease ($\approx$ 11\%/19\%) in the TMR of the Sk/u-MTJ as temperature increases from $0-450\textdegree$ K. To understand this roll-off, we analyze the transmission spectra as shown in Fig.\ref{fig:IV}(d). The TMR, expressed in terms of the transmission coefficient, is given by
\begin{equation}
TMR= \frac{\int {T_D(E)}(f_T-f_B)dE}{\int T_{AP}(E)(f_T-f_B)dE} \times 100
\label{eqn:TMR_2} 
\end{equation}
where $f_{T/B}$ is the Fermi-Dirac distribution function of the top/bottom contact. As the temperature increases, the fermi-window ($FW = f_T - f_B$) below and above the fermi energy broadens. Initially, the contribution to the numerator of Eq.~\ref{eqn:TMR_2} remains nearly constant because the FW below and above the fermi energy encompasses an increasing and decreasing transmission difference ($T_D = T_{AP} - T_P$), respectively. This leads to a gradual roll-off in the TMR (see Fig.~\ref{fig:IV}(c) \& (d)). At higher temperatures, as the FW crosses the energy point corresponding to the $T_D(E)$ peak ($2.13, \text{eV}$ in this case), the additional contribution to the transmission difference from the broader FW begins to decrease. This results in a more pronounced TMR roll-off effect at higher temperatures. In a similar manner, the TMR roll-off with temperature in u-MTJs can be rationalized even without invoking magnon scattering to the first order, in contrast to the earlier works\cite{Temp1, Temp2, Temp3}.\\
\begin{figure}
    \centering
    \includegraphics[width=\linewidth]{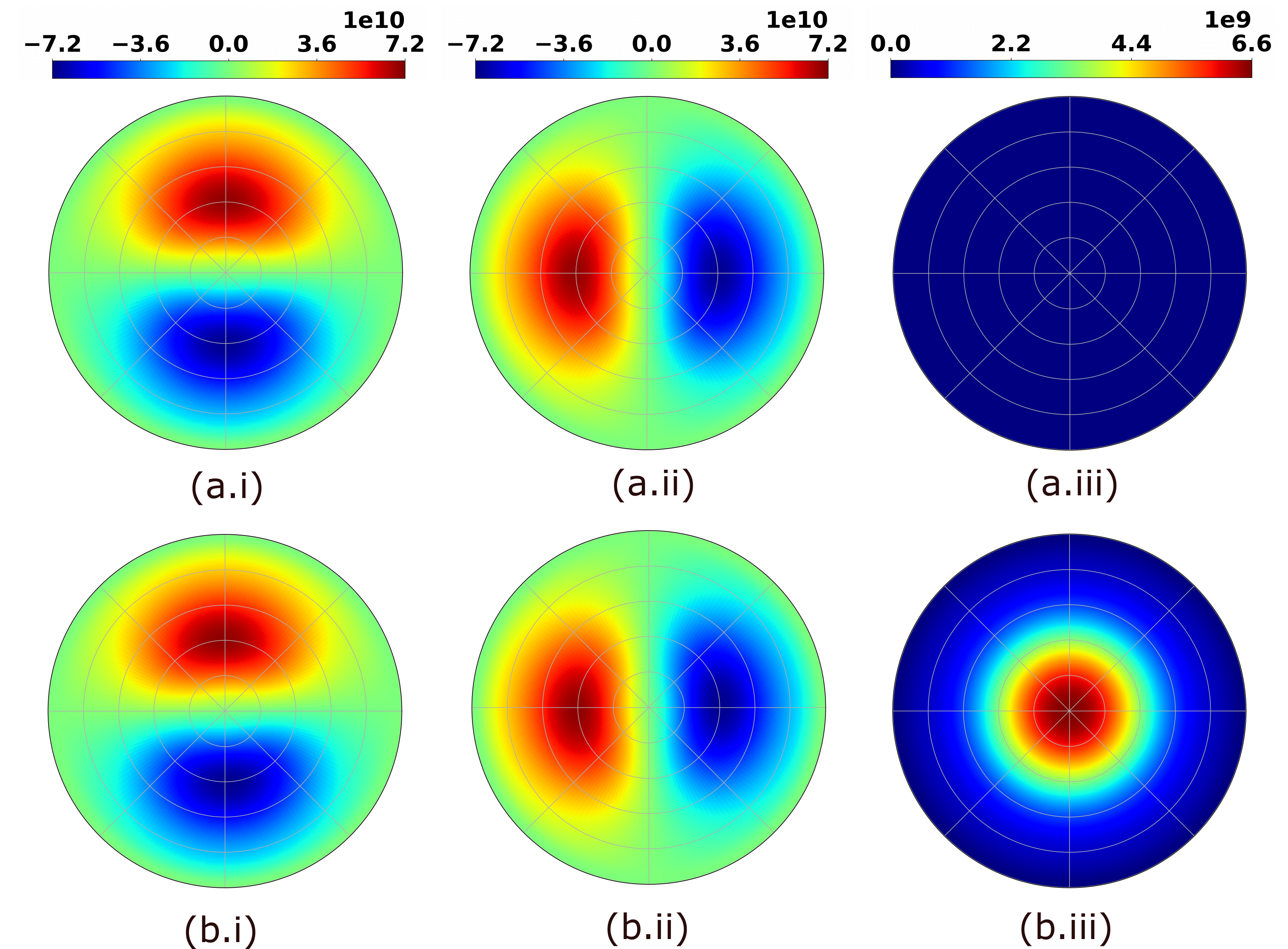}
  \caption{Spin current density($A/m^2$) (i) $I_x$, (ii) $I_y$, (iii) $I_z$ at (a) zero volt bias and at (b) 10 mV bias for Neel-type skyrmion on the Sk-MTJ in the PC.}
  \label{fig:SNeel}
\end{figure}
\begin{figure}
  \centering
    \centering
    \includegraphics[width=\linewidth]{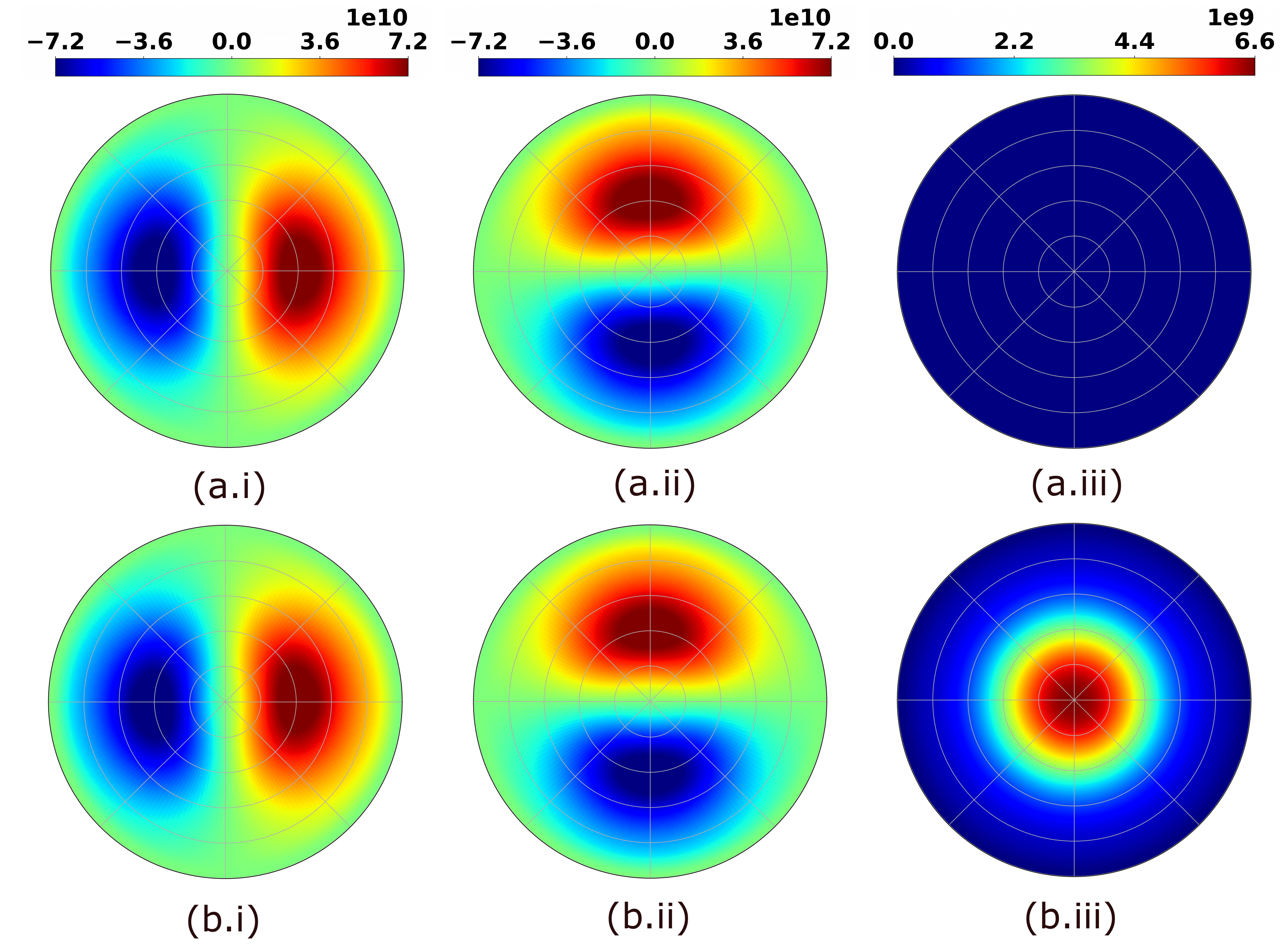}
  \caption{Spin current density($A/m^2$) (i) $I_x$, (ii) $I_y$, (iii) $I_z$ at (a) zero volt bias and at (b) 10mV bias for Bloch-type skyrmion on the Sk-MTJ in the PC.}
  \label{fig:SBloch}
\end{figure}
\noindent The charge current density of the Sk-MTJ depends on the azimuthal projection of the skyrmion texture onto the u-FM. As a result, the Sk-MTJs with Néel- and Bloch-type skyrmions exhibit identical current density profiles as shown in Fig.~\ref{fig:Charge}(a.i, a.ii) \& (b.i, b.ii), in the collinear configuration of u-FM and skyrmion core. In the PC, the current density is highest at the center ($\approx 8 \times 10^9  \text{A/m}^2$) and gradually decreases radially outward. In contrast, in the APC, the current density is significantly lower at the center ($\approx 2 \times 10^9 \text{A/m}^2$), increases radially outward to a peak of $\approx 8 \times 10^9 \text{A/m}^2$, and then reduces to zero at the edges. The circular symmetry in the charge current density distribution for both the PC and the APC arises from the inherent circular symmetry of the skyrmion and the u-FM. However, the circular symmetry in the current density profile disappears in the non-collinear configuration of the u-FM and the skyrmion core as shown in as shown in Fig.~\ref{fig:Charge}(a.iii) \& \ref{fig:Charge}(b.iii). Consequently, local charge current measurements can be used to distinguish between Néel-type and Bloch-type skyrmion on MTJ.\\
\begin{figure}
  \centering
    \includegraphics[width=\linewidth]{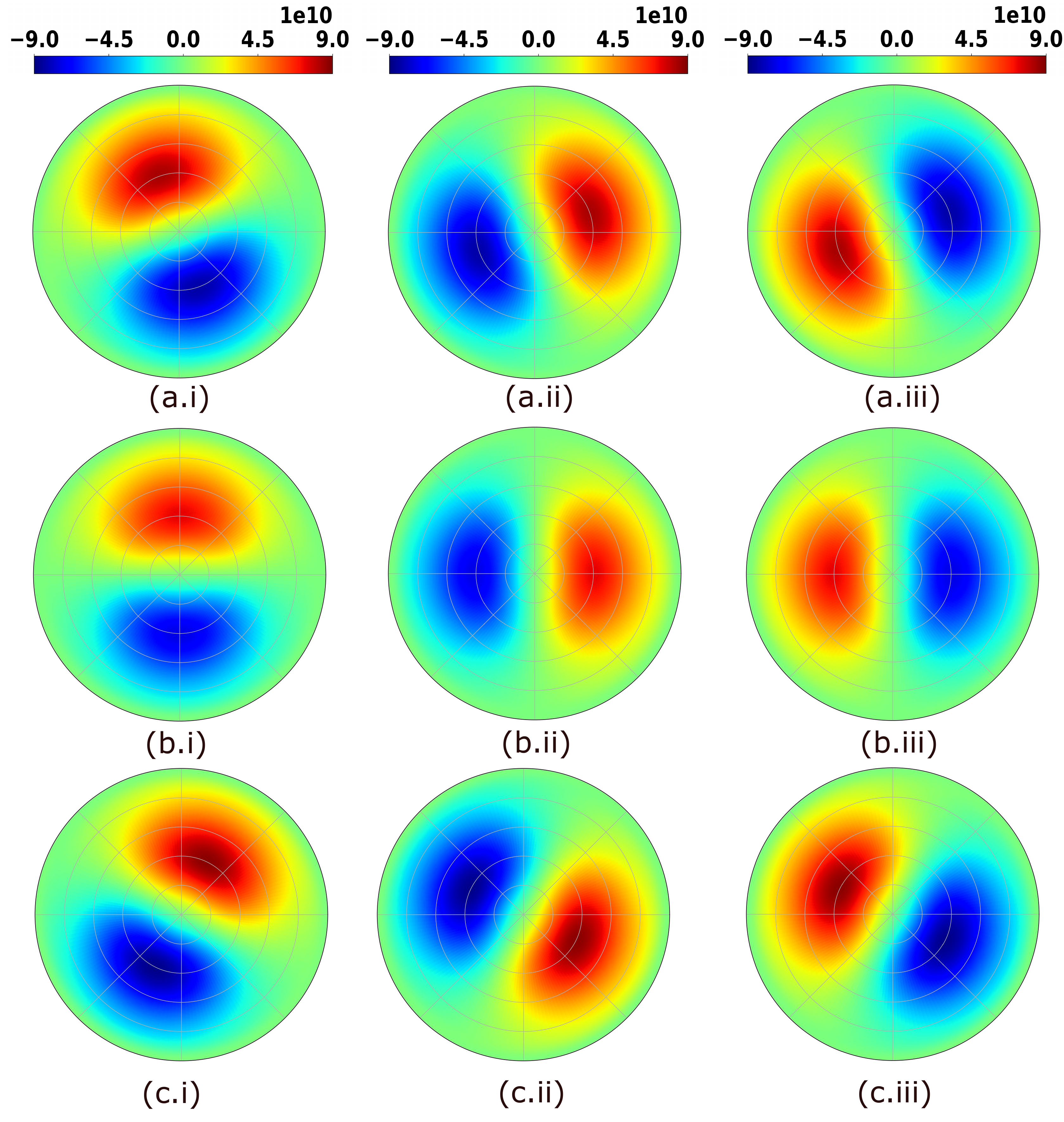}
  \caption{Spin current($A/m^2$) $I_x$ emerging from the u-FM when Sk-MTJ have (i) Néel (ii) right-handed Bloch and (iii) left-handed Bloch type skyrmion at a bias voltage of a) 0.1 V, b) 0 V and c) -0.1 V in the PC.}
  \label{fig:Rot}
\end{figure}
\indent The spin current emerging from the u-FM in the Sk-MTJ exhibits a textured spatial profile. Figures~\ref{fig:SNeel}(a.i, a.ii) and \ref{fig:SBloch}(a.i, a.ii) illustrate the emergence of a spatially dependent exchange coupling \cite{SC1}, also referred to as dissipation-less spin currents ($I_x$ \& $I_y$) at zero bias in the Sk-MTJ. At non-zero bias, the $I_z$ spin current density emerging from the u-FM remains identical for Néel- and Bloch-type skyrmions in the Sk-MTJ, as shown in Fig.~\ref{fig:SNeel}(b.iii) \& \ref{fig:SBloch}(b.iii), due to circular symmetry in collinear configuration. However, the spin current densities $I_x$ and $I_y$ are rotated by $\frac{\pi}{2}$ between Néel- and Bloch-type Sk-MTJs, as illustrated in Fig.~\ref{fig:SNeel}(b.i, b.ii) and Fig.~\ref{fig:SBloch}(b.i, b.ii). The net spin current emerging from the u-FM in the Sk-MTJ exhibits helicity for Néel-type skyrmion but lacks the same for Bloch-type skyrmion. Figure~\ref{fig:Rot} shows spacial rotation of exchange coupling strength ($I_x$, a similar behavior is observed for $I_y$) with the voltage variation ($-0.1 \, \mathrm{V}$ -- $0.1 \, \mathrm{V}$) in the collinear configuration. The direction and angle of rotation are influenced by factors such as the u-FM configuration and the skyrmion texture. This behavior not only enables the identification of different skyrmion types but also reveals unique characteristics, such as helicity in Neel-type skyrmion, through local measurements of spin current emanating from the u-FM in Sk-MTJ at various voltages.\\
\begin{figure}
    \centering
    \includegraphics[width=\linewidth]{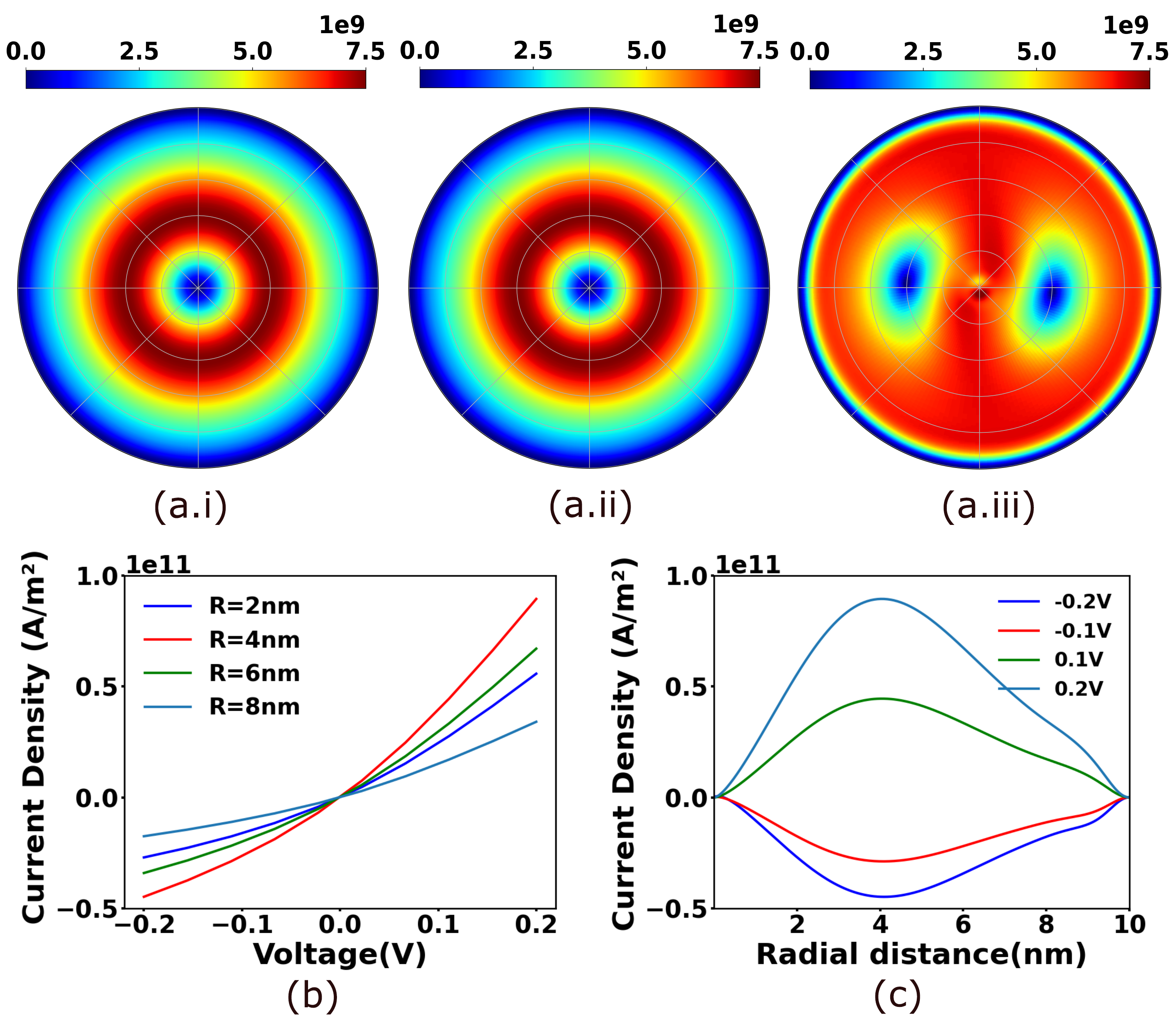}
  \caption{Damping-like term ($I_\tau$ in a.u) with spatial variation at 20 mV for (a.i) the PC, (a.ii) the APC and (a.iii) orthogonal configuration. (b) Voltage variation of $I_\tau$ at different radial distances and (c) radial variation of $I_\tau$ at different voltages in the PC.}
\label{fig:damp}
\end{figure}
\begin{figure}
    \centering
    \includegraphics[width=\linewidth]{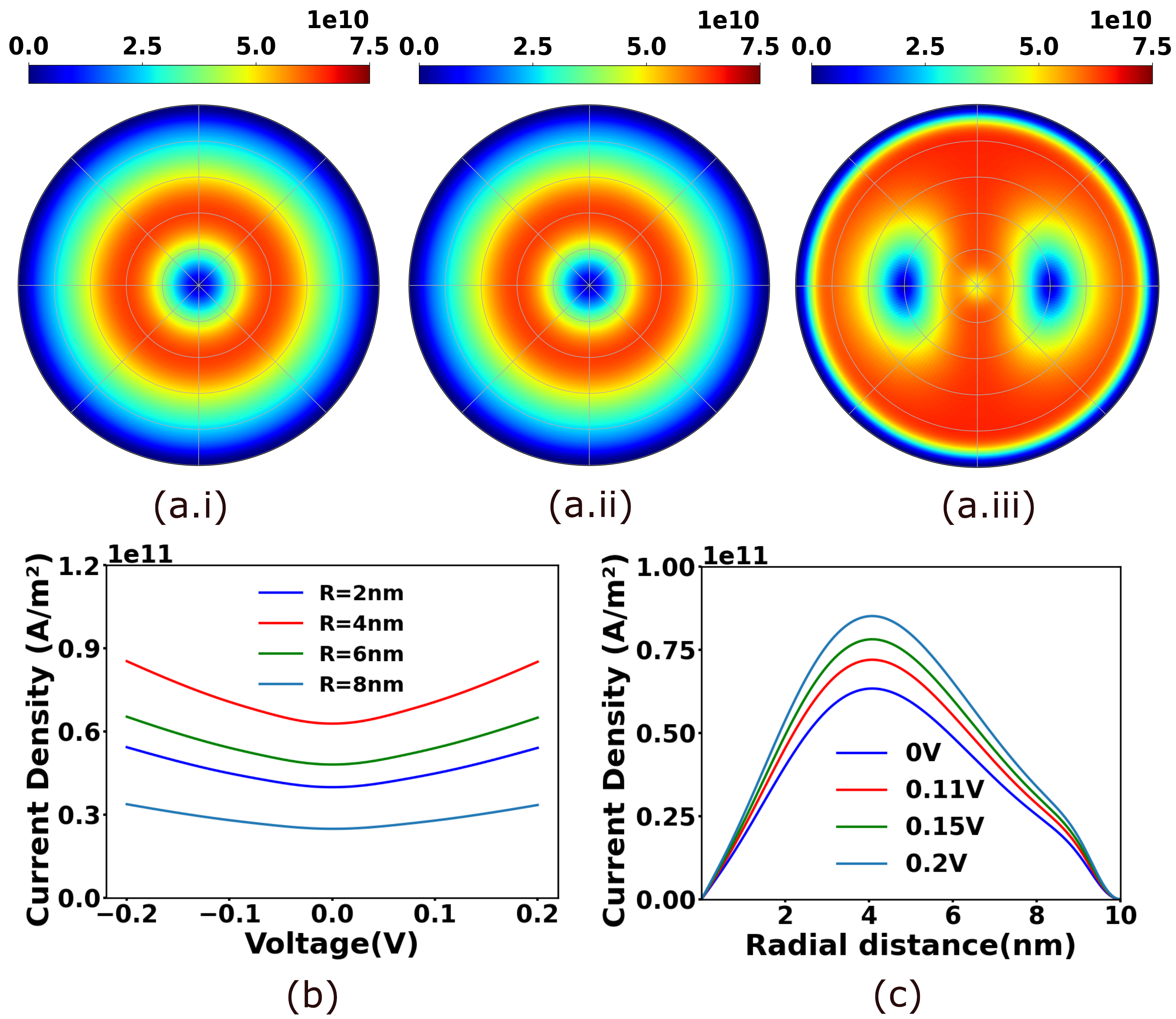}

  \caption{Field-like term ($I_h$ in a.u) with spatial variation at 20 mV for (a.i) the PC, (a.ii) the APC and (a.iii) orthogonal configuration. (b) Voltage variation of $I_h$ at different radial distances and (c) radial variation of $I_h$ at different voltages in the PC.}
\label{fig:field}
\end{figure}
\begin{figure}
  \centering
    \includegraphics[width=\linewidth]{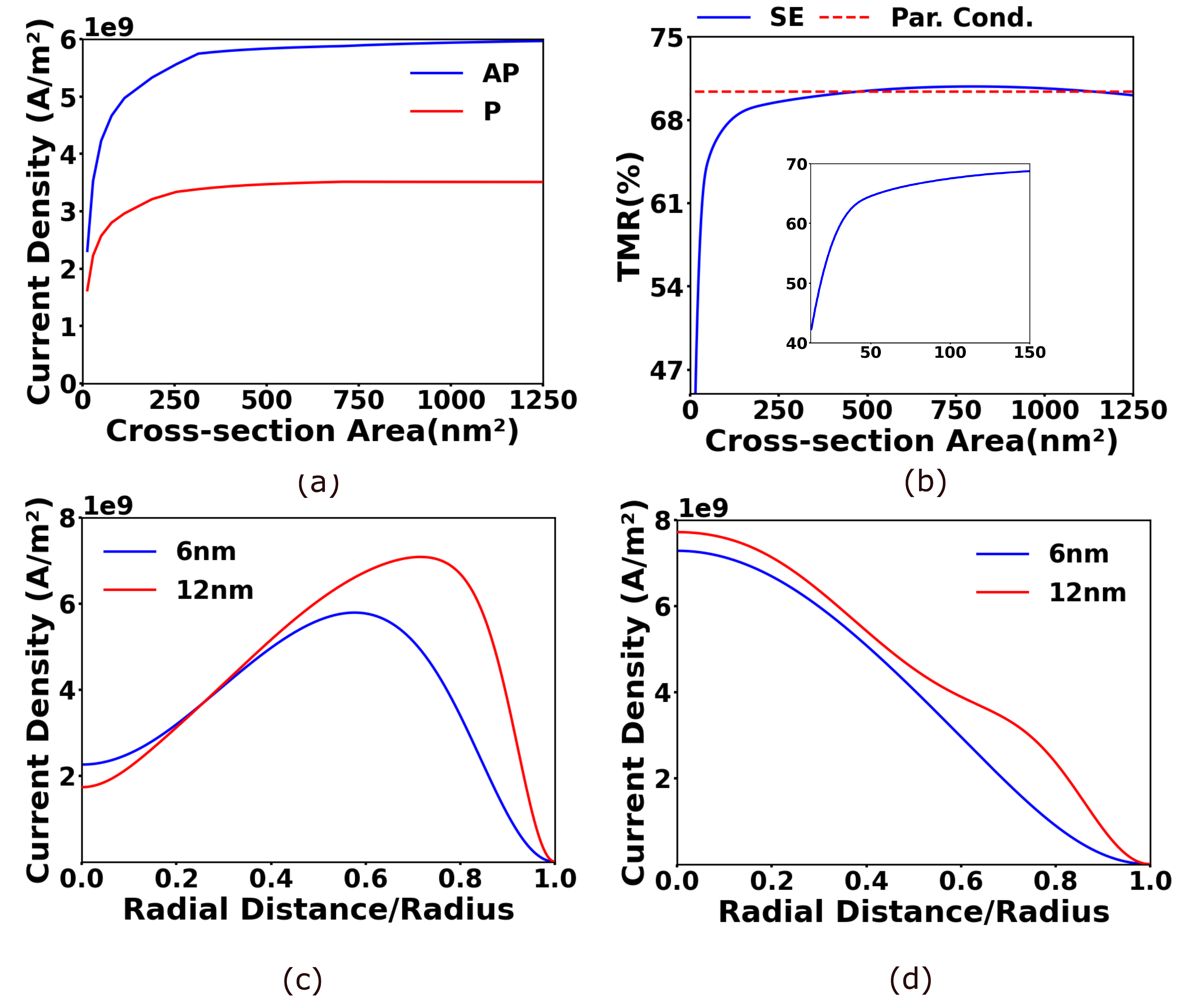}
  \caption{ (a) Variation of charge current and (b) TMR with the cross-section area of the Sk-MTJ calculated using spatio-eigen (SE) and parallel conductance (Par. Cond.) approach. The inset shows a zoomed-in version of the TMR roll-off for smaller cross-section areas. Radial variation of charge current in the Sk-MTJ with a cross-sectional diameter of 12nm and 6nm in (c) the APC and (d) the PC at 10mV.}
\label{fig:size}
\end{figure}
\begin{figure}
  \centering
    \includegraphics[width=\linewidth]{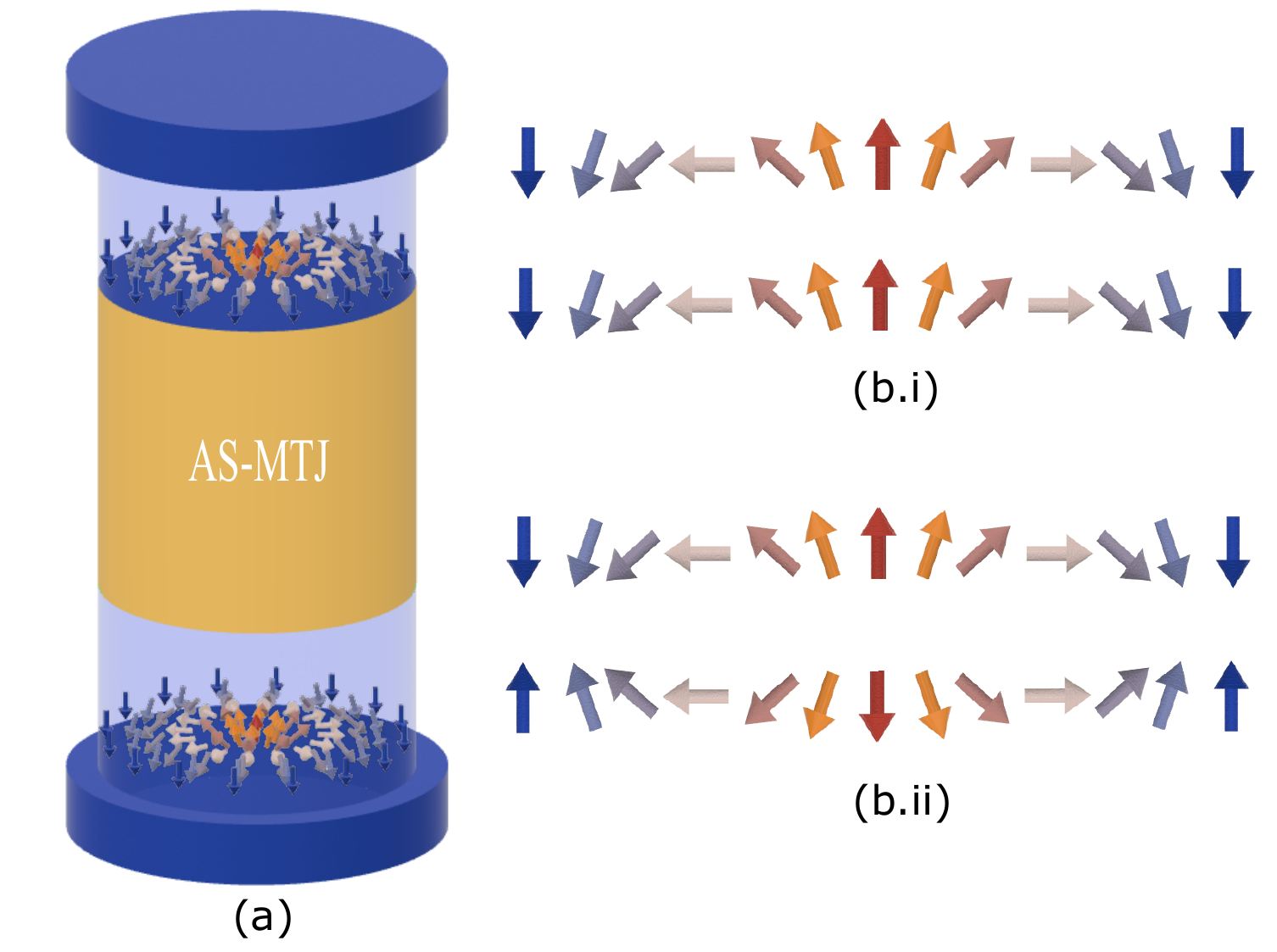}
  \caption{ (a) Schematics of AS-MTJ. (b) The relative orientation of top skyrmion and bottom skyrmion in (i) the PC and (ii) APC.}
\label{fig:SkyAll}
\end{figure}

An alternative perspective on spin current is particularly relevant when investigating the torque induced by spin current. The non-collinearity between the magnetization direction of the u-FM ($\hat{M}$), and the magnetization of the skyrmion layer ($\hat{m}(\vec{r})$), enables the u-FM layer to induce dynamics in the skyrmion layer via spin-transfer torque (STT)\cite{STT,SkySST1,SkySST2}. In this scenario, the spin current can be expressed as 
\begin{eqnarray}
    \vec{I_s}(r) = I_{0}(r)\hat{m} +I_{h}(r)\hat{m}_h+I_{\tau}(r)\hat{m}_{\tau}
\end{eqnarray}

here, \(\hat{m}_{h}\) and \(\hat{m}_{\tau}\) are unit vectors along \(\hat{m}(r) \times \hat{M}\) and \(\hat{m}(r) \times (\hat{m}(r) \times \hat{M})\), respectively. In this decomposition, \(I_{\tau}\) and \(I_{h}\) are referred to as the damping/anti-damping (Slonczewski) term and the field-like term, respectively. In the Sk-MTJ, the \(I_{\tau}\) term, which represents the damping/anti-damping effect exerted by the u-FM on the skyrmion-textured layer, exhibits radial variation along with circular symmetry in the PC/APC configuration, as shown in Fig.~\ref{fig:damp}(a). Consequently, different regions of the skyrmion experience varying torques, which can significantly impact various skyrmion-based applications, including skyrmion switching in MTJs \cite{MTJ2024} and skyrmion-based quantum gate operations \cite{SkyQ2}. The circular symmetry of \(I_{\tau}\) can be lifted in the Sk-MTJ when the u-FM is magnetized in-plane, as illustrated in Fig.~\ref{fig:damp}(a,iii). As shown in Fig.~\ref{fig:damp}(b) and (c), \(I_{\tau}\) exhibits an asymmetric voltage profile, where both the voltage polarity and the degree of asymmetry depend on the radial distance. This bias-dependent asymmetry and radial variation in the profile result from the combined effects of skyrmion micro-magnetic orientations and boundary conditions. 

The \(I_h\) term, which is associated with exchange coupling and field-like torque, retains a circularly symmetric pattern in the PC/APC configuration, as depicted in Fig.~\ref{fig:field}. However, unlike the damping/anti-damping torque, the field-like torque does not exhibit asymmetry for positive and negative voltages. Nevertheless, this component has a nonzero value at zero bias, which varies radially \cite{SC1}, signifying the FM exchange.\\
\indent The scaling of MTJs and skyrmions plays a crucial role in driving technological miniaturization and enabling energy-efficient applications\cite{MTJS1,MTJS2}. In this section, we investigate the impact of scaling on the Sk-MTJ. We also contrast the scaling results obtained using the proposed coupled spatio-eigen NEGF framework with earlier proposed effort\cite{Qt_um_2022}.

Figure~\ref{fig:size}(a) illustrates that the charge current density remains independent of the cross-sectional area, as expected, for Sk-MTJs with larger dimensions. However, at smaller scales, the pronounced discretization of transverse energy levels ($\epsilon_i$) results in a reduction in charge current density. At lower cross-sectional areas, we observe a roll-off in the TMR for the Sk-MTJ. In previous attempts to model such systems with micromagnetic contacts \cite{Qt_um_2022}, an effective parallel conductance model was proposed. In P. Flauge et al.'s \cite{Qt_um_2022} approach, the current density in each domain is calculated using a 1D NEGF, and the total current profile is subsequently assembled by stitching together the contributions from each domain. However, in reality, these domains are interconnected, as represented by the off-diagonal terms in the tight-binding model of the transverse Hamiltonian (Eq.~\ref{eq:HamilF_x}). At smaller scales, the effects of this cross-coupling become pronounced, leading to a clear deviation from the effective parallel conductance model. This deviation results in a roll-off in TMR, whereas the parallel conductance model predicts the TMR to remain unaffected, as shown in Fig.~\ref{fig:size}(b). The reduction in the Sk-MTJ's TMR becomes more pronounced as the device diameter shrinks below 10 nm, indicating strong interference between the different nano-domains of the skyrmion in the Sk-MTJ along with the boundary effect. To understand the TMR decrease with reduced area in the Sk-MTJ, we examine the current density at diameters of 12 nm and 6 nm in the PC and the APC (Fig.~\ref{fig:size}(c,d)). For the PC, there is an overall reduction in current density, with a particularly pronounced effect near the boundaries. In the APC, the maximum current density not only decreases (by approximately 18\%) but also shifts horizontally. The reduction in current density at smaller diameters can be attributed to the dominance of boundary effects. However, the horizontal shift in peak current density(Fig.~\ref{fig:size}(c)), coupled with an approximately 30\% increase at the center for the 6 nm Sk-MTJ, can be explained only by the interference between different nano-domains within the skyrmion texture. The coupled transverse eigenchannels of the transverse Hamiltonian (see Eq. \ref{eq:HamilF_x}) in the Sk-MTJ translates the interference between nano-domains as the coupling between regions of higher and lower conduction (see Fig. \ref{fig:NonUnifrom}). This interaction enables lower-conduction pathways to tunnel through alternative routes, effectively enhancing the overall conductance. As a result, the interference reduces the transmission disparity between the PC and the APC, ultimately causing a decrease in TMR at smaller areas of the Sk-MTJ.\\
\begin{figure}
  \centering
    \includegraphics[width=\linewidth]{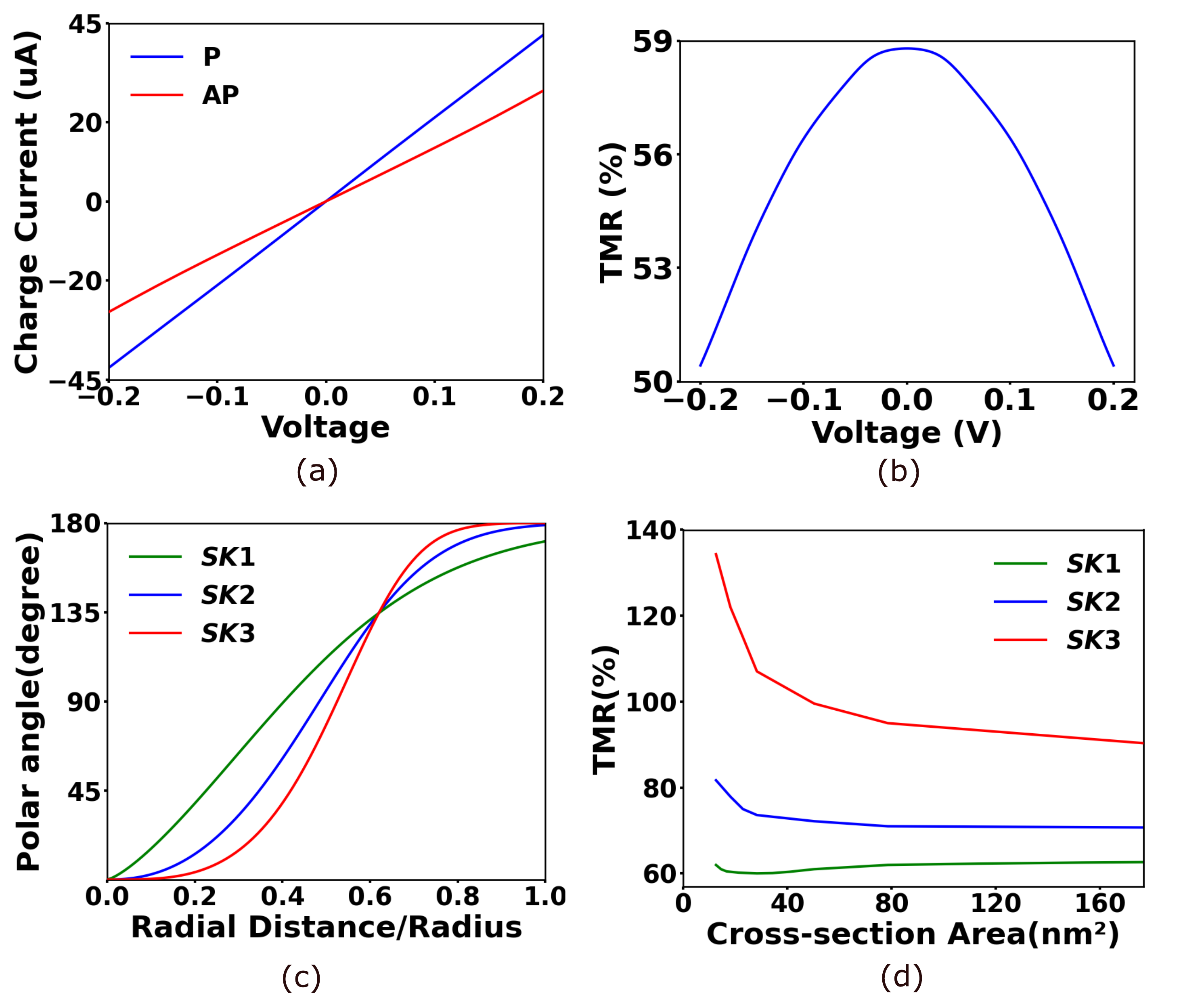}
  \caption{(a) Current-Voltage characteristics in the PC and the APC, (b) TMR variation with voltage for the AS-MTJ. (c) Variation of the direction of magnetization over the radial profile for different skyrmions, (d) TMR variation with cross-sectional area of the AS-MTJ for different skyrmions at 10mV.}
\label{fig:AllSky}
\end{figure}
\begin{figure}
  \centering
    \includegraphics[width=\linewidth]{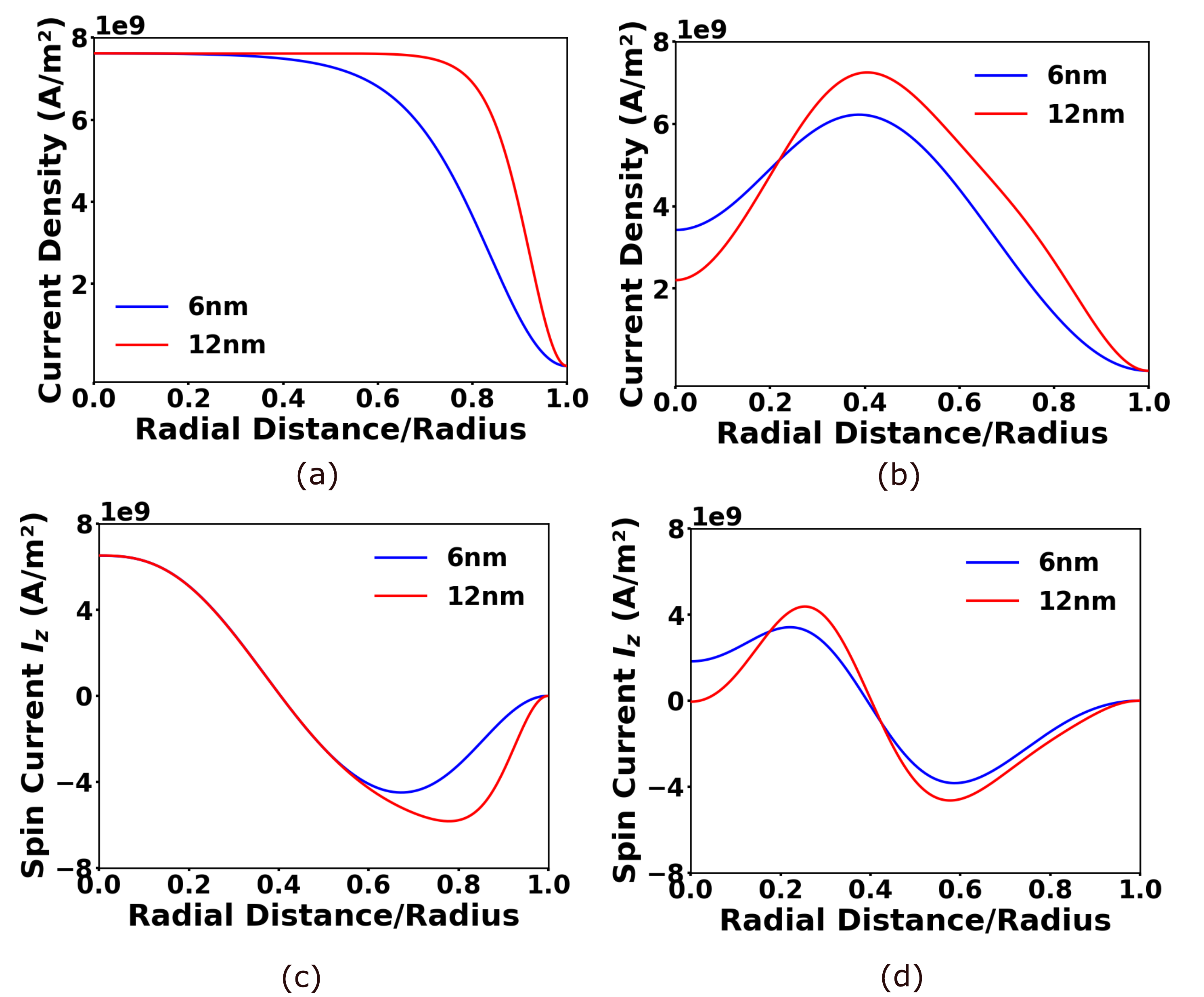}
  \caption{Radial variation of charge current in (a) the PC, (b) the APC, and the spin current ($I_z$) in (c) the PC, (d) the APC for the AS-MTJ with a cross-sectional diameter of 12nm and 6nm, at 10mV bias.}
\label{fig:Allsky126}
\end{figure}
Finally, we explore the impact of scaling in the all-skyrmion-MTJ (AS-MTJ), where both layers exhibit magnetic skyrmion textures. The parallel and anti-parallel configurations (PC \& APC) (Fig.~\ref{fig:SkyAll}(b)) are distinguished by the relative alignment of the skyrmion centers. We describe the scaling impact on the TMR, the charge, and spin currents. We also analyze the influence of scaling on AS-MTJs with skyrmions of varying domain wall widths\cite{Wall}, which represent the effective distance beyond the skyrmion diameter over which the magnetization transition occurs (Fig.~\ref{fig:AllSky}(c)). We have considered three skyrmions—$Sk_1$, $Sk_2$, and $Sk_3$—each characterized by progressively narrower domain walls with the same diameter. For most of the discussion on spin and charge current unless otherwise specified, the discussion primarily focuses on the $Sk_1$ Néel-type skyrmion.

We show in Fig.~\ref{fig:AllSky}(a) the IV characteristics of the AS-MTJ having the same skyrmion in both layers. The PC of the AS-MTJ exhibits currents comparable to the PC of the u-MTJ. However, the TMR of the AS-MTJ is lower than that of both the u-MTJ and Sk-MTJ (see Fig. \ref{fig:IV}(b)), decreases with voltage in an expected manner as shown in Fig.~\ref{fig:AllSky}(b). Notably, the TMR roll-off behavior is influenced by the skyrmion's domain configuration. In the AS-MTJ with the $Sk_1$ skyrmion (Fig.~\ref{fig:AllSky}(d)), the TMR initially exhibits a slight roll-off as the cross-sectional area decreases and, a minor increase with further reduction in area. This behavior contrasts significantly with that of the Sk-MTJ (see Fig.\ref{fig:size}(b)). As the domain wall becomes thinner, the TMR no longer exhibits a dip; instead, it shows a gradual increase as the cross-sectional area decreases. With further reduction in domain wall thickness, the TMR rises significantly with decreasing cross-sectional area. 
We can refer to the radial current profile of the AS-MTJ with $Sk_1$ skyrmion to explain these phenomena(see Fig.~\ref{fig:Allsky126}(a,b)). Since we use skyrmions with the same radius and domain size, the layers are effectively aligned across most locations in the PC, resulting in a uniform charge current across the cross-section, except for diminished current near the boundaries(Fig.~\ref{fig:Allsky126}(a)). In the APC, we observe a low-conductance region at the center and edges—where the FM layers are effectively anti-parallel and a higher conductance region in between. As domain wall width is reduced (Fig.~\ref{fig:Allsky126}(c)), the current profile is expected to remain unchanged in the PC. However, in the APC, the high-conductance peak becomes narrower, leading to a reduction in the overall charge current and an increase in TMR, as observed in Fig.\ref{fig:AllSky}(d). 

When the device is scaled down from 12 nm to 6 nm, the PC shows no significant change, apart from boundary effects that reduce conduction in deeper regions, leading to an overall decrease in conductance at a small cross-section (Fig.~\ref{fig:AllSky}(c)). In the APC, the boundary effect reduces the maximum of the high-conductance region. Hence, we expect a greater reduction in the fraction of charge current in the APC compared to the PC. However, quantum tunneling increases the current density near the center due to the cross-coupling of transverse eigenchannels (see Fig.~\ref{fig:NonUnifrom}). The interplay of the opposing effects of boundary and cross-coupling of transverse eigenchannels determine the small-scale behavior of TMR. For $Sk_1$ (Fig.~\ref{fig:AllSky}(d)), this interplay initially causes a slight reduction in TMR. As the boundary effect intensifies at smaller cross-sections, the TMR eventually exhibits a slight increase. For a thinner domain wall width, TMR increases significantly as the skyrmion size decreases. This occurs because, for a thinner wall width in the APC configuration, the high-current region becomes localized near the outer edges. As a result, quantum tunneling through the low-conductance central region is reduced, allowing boundary effects to dominate. In a nutshell, the TMR variation with area reduction is a complex interplay of boundary effect and quantum tunneling through connected transverse eigenchannels, leading to varying behaviors across the various possible Sk/AS-MTJ devices. These effects highlight the importance of exploring different device configurations to identify those that maximize TMR while maintaining scalability as device sizes decrease.\\
\begin{figure}
    \centering
    \includegraphics[width=\linewidth]{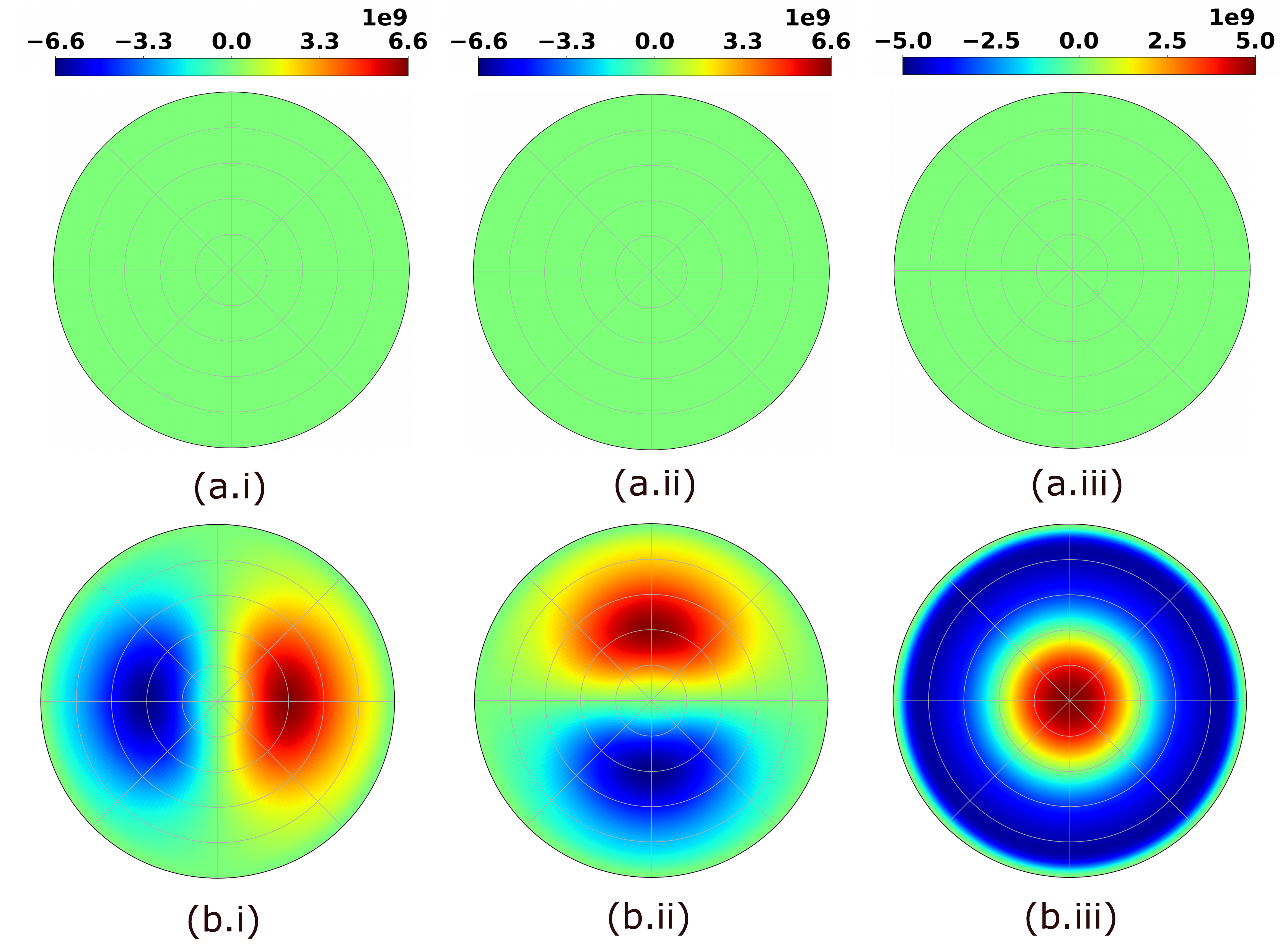}
  \caption{Spin current density($A/m^2$) (i) $I_x$, (ii) $I_y$, (iii) $I_z$ at (a) zero volt bias and at (b) 10mV bias for the AS-MTJ in the PC.}
  \label{fig:ASSpinU}
\end{figure}
\begin{figure}
  \centering
    \centering
    \includegraphics[width=\linewidth]{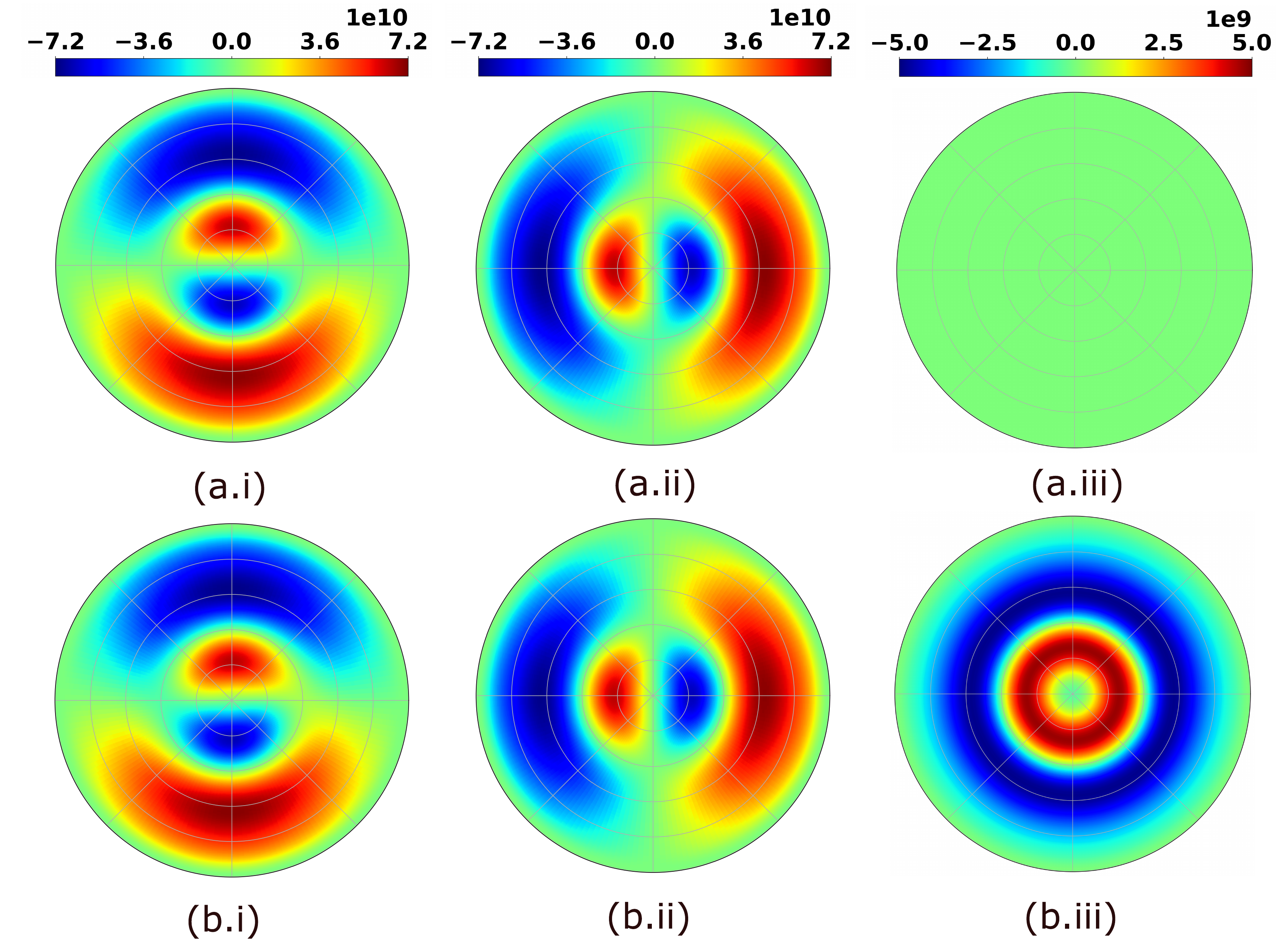}
  \caption{Spin current density($A/m^2$) (i) $I_x$, (ii) $I_y$, (iii) $I_z$ at (a) zero volt bias and at (b) 10mV bias for the AS-MTJ in the APC.}
  \label{fig:ASSpinN}
\end{figure}

\indent Extending our exploration, we now examine the spin current profile of the AS-MTJ. In the PC, there is no spin current at zero bias in any direction due to the aligned magnetization of both FM layers. However, at a finite bias (10 mV), the spin current reaches the order of \( 5 \times 10^9 \, A/m^2 \). In the APC, there is non-zero spin exchange coupling \cite{SC1} current in the \(x\)- and \(y\)-directions at zero bias, and these directions dominate at small bias, while the \(z\)-direction still exhibits zero spin current at zero bias.

At first glance, the spin current pattern appears to follow the relative local orientation of the FM domains, especially in the PC. However, interconnections between domains (represented by tight-binding in the \(xy\)-direction) can introduce intriguing quantum effects, particularly as the device’s cross-section is scaled down. As shown in Fig.~\ref{fig:Allsky126}(c), in the PC, the primary boundary effect causes a faster decay of spin current density at the edges. In contrast, intriguing patterns emerge in the APC. For larger devices (12 nm diameter, Fig.~\ref{fig:Allsky126}(d)), the \(z\)-component of the spin current at the center is nearly zero due to the anti-parallel alignment of the domains, leading to equal tunneling of both up and down spins. However, as we shrink the device to a 6 nm diameter (Fig.~\ref{fig:Allsky126}(d)), a non-zero spin current appears at the center, likely due to electron tunneling from other domains. The scaling study of ST-MTJ and AS-MTJ suggests that device behavior—including TMR, spin current, and other characteristics—varies significantly as the device size decreases. To accurately model these effects, it is essential to employ comprehensive models, such as the one we propose, to account for all relevant interactions.

\section{Conclusion}
In this work, we delineate the charge and spin characteristics of the Sk-MTJ with both Bloch and Néel-type textures across various voltages, temperatures, and sizes. We detail the emergence of a textured spin current originating from the uniform layer of Sk-MTJs and outline a radially varying, asymmetrical voltage dependence of spin torque. Furthermore, we describe the voltage-induced rotation of the spin current texture, coupled with the emergence of helicity in the spin current, particularly evident for Néel skyrmions in MTJs under non-equilibrium conditions. We identify the temperature roll-off in the TMR of Sk-MTJs and u-MTJs, attributing it to ballistic transmission spectra, abating the need for magnons to the first order. Additionally, we describe the effect of scaling on the Sk-MTJs and explain the observed TMR roll-off based on the charge current profile.\\
\indent We also analyze the AS-MTJ device, focusing on its IV characteristics and TMR, as well as the effects of scaling and variations in skyrmion parameters. Notably, while the TMR of Sk-MTJs decreases with a reduction in diameter, the TMR of AS-MTJs remains constant or exhibits an increase as the diameter decreases. This increase is closely tied to the domain wall of the skyrmion. Furthermore, we examine the charge and spin currents in the AS-MTJ and investigate how scaling impacts these properties. Our results indicate that the spin current profile undergoes significant changes as the device size decreases, which can, in turn, influence magnetic dynamics through spin-transfer torque.\\
\indent These findings underscore the importance of skyrmion properties—such as diameter and domain-wall size—as well as device dimensions in the design of the Sk/AS-MTJs for next-generation memory and spintronic devices. This work provides a pathway for optimizing TMR and spin current in conjunction with device scaling. Moreover, we present a new formalism offering a generic and computationally efficient approach to integrate micromagnetic simulations with quantum transport calculations, enabling the exploration of diverse topologically non-trivial magnetic phases in MTJs \cite{Beyond}.

\appendix
\section{Current Operator}
\label{CurrentOp}
In this section, we show that similar to the transmission operator, the current operator can also be reduced to the relevant terms plus some terms with $u^2$ order. Recall that the current operator is given by 
\begin{equation}
\hat{\Pi}(E) =\frac{i}{\hbar} \left( H_{N-1,N} G^n_{N,N-1} - \left( H_{N,N-1} G^n_{N-1,N} \right)^\dagger \right)
\label{eq:IopA}
\end{equation}
here, $H$ is Hamiltonian of the device and $G^n$ is less than Green's function, define as
\begin{equation}
    G^n = i G \, \Gamma_B \,G^{\dagger}\,f_B + i G \, \Gamma_T \, G^\dagger \,f_T
\end{equation}
Plugging the expression for the lesser Green's function into the current operator yields four terms. Let’s select one term, expressed as
\begin{equation}
\hat{\Pi}_1 =\frac{t_0}{\hbar} \left(G_{N,1}\Gamma_BG^{\dagger}_{N-1,1}\right)f_B(E)
\label{eq:IopA_1}
\end{equation}
here, we used $H_{N-1,N} = -t_0 \mathbf{I}$. Expressing Eq.\ref{eq:IopA_1} in the form of the relevant and irrelevant terms, starting from the broadening matrix as
\begin{equation}
\begin{array}{l}
\Gamma_{B} =\left(\begin{array}{ll}
\Gamma_{11} & 0 \\
0 & 0
\end{array}\right)_{\lambda}
\end{array}
\end{equation}
Similarly, the matrix $G_{N,1}$ and $G_{N-1,1}$ can be represented in a $2 \times 2$ block matrix as
\begin{equation}
\begin{array}{cc}
G_{N,1} = \left(\begin{array}{ll}
g^a_{11} & g^a_{12} \\
g^a_{21} & g^a_{22}
\end{array}\right)_{\lambda}, &
G_{N-1,1} = \left(\begin{array}{ll}
g^b_{11} & g^b_{12} \\
g^b_{21} & g^b_{22}
\end{array}\right)_{\lambda}.
\end{array}
\end{equation}
Since the spin and charge current calculation involve a trace (Eq.~\ref{eq:Ic} \&~\ref{eq:Is}), terms of the current operator effectively reduce to:
\begin{equation}
\hat{\Pi}_{1,eff} =\frac{t_0}{\hbar} \left(g^a_{11}\Gamma_{11}g^{b,\dagger}_{11}+g^a_{21}\Gamma_{11}g^{b,\dagger}_{21}\right)f_B(E)
\label{eq:IopA3}
\end{equation}
Now, we can follow a similar procedure, as used in the methodology section, to reduce the transmission matrix in terms of the relevant terms for the various parts of the current operator. Using the Eq. \ref{eq:Grr} and Eq. \ref{eq:ABC}, one can show that:
\begin{subequations}
    \begin{equation}
    G_{N,1}=  -t_m(D_{N-1,1}^{-1})(A-t_m^{2}(D_{1,1})^{-1})^{-1}
    \end{equation}
    \begin{equation}
    G_{N-1,1}=  -t_m(D_{N-2,1}^{-1})(A-t_m^{2}(D_{1,1})^{-1})^{-1}
    \end{equation}
\end{subequations}
As we expand the above expression into relevant and irrelevant terms (Eq.~\ref{eq:AD}) and substitute it into Eq.~\ref{eq:IopA3}, we can show that the first term ($g^a_{11}\Gamma_{11}g^{b,\dagger}_{11}$) reduces to just the relevant terms, similar to the transmission operator, and \( g^a_{21} \Gamma_{11} g^{b,\dagger}_{21} \approx O(u^2) \), i.e., of order \( u^2 \) or higher. In fact, the procedure can be repeated for the remaining terms in Eq.~\ref{eq:IopA} to show that they also reduce to the relevant part plus some terms of order \( u^2 \) or higher. As discussed at the end of the methodology section, the matrix \( u \) characterizes the coupling between the higher relevant eigen-channels and the lower-order irrelevant eigen-channels. This factor decreases as the relevance threshold increases, thereby reducing its impact. Consequently, the current expression can be constructed using only the relevant terms with an appropriate relevance threshold ($E_t$).

\section{Toy model}
\label{appA}
We demonstrate the effectiveness of the spatio-eigen approach within the NEGF framework using a 2D magnetic tunnel junction (MTJ)-like structure, where a 2D insulator is positioned between two 2D ferromagnetic (FM) contacts. One contact exhibits uniform magnetization, while the other features a non-uniform magnetization texture, with the in-plane angle of magnetization varying according to a Gaussian distribution.\\
\indent The Hamiltonian for this model is similar to that described in Eq. (\ref{eq:Hamill}), except that the transverse Hamiltonian is given by
\begin{equation}
   H_{B/T}^{FM} = (H^{FM}_x + 2\operatorname{t_m}\mathbf{I}_x) \otimes \mathbf{I}_2 + \Delta_{B/T.} 
\label{eq:Ap1}
\end{equation}
Similarly, the tight-binding term along the y-axis is omitted in the remaining transverse Hamiltonian (Eig. \ref{eq:Hamil_c}, \ref{eq:Hamil_I}). The rest of the parameters we will keep the same as in Sec.\ref{sec:Model}. Hence, the transverse Hamiltonian will be of size $(800 \times 800)$ for a 2D-MTJ with a width of 10 nm.

\begin{figure}
    \centering
    \includegraphics[width=\linewidth]{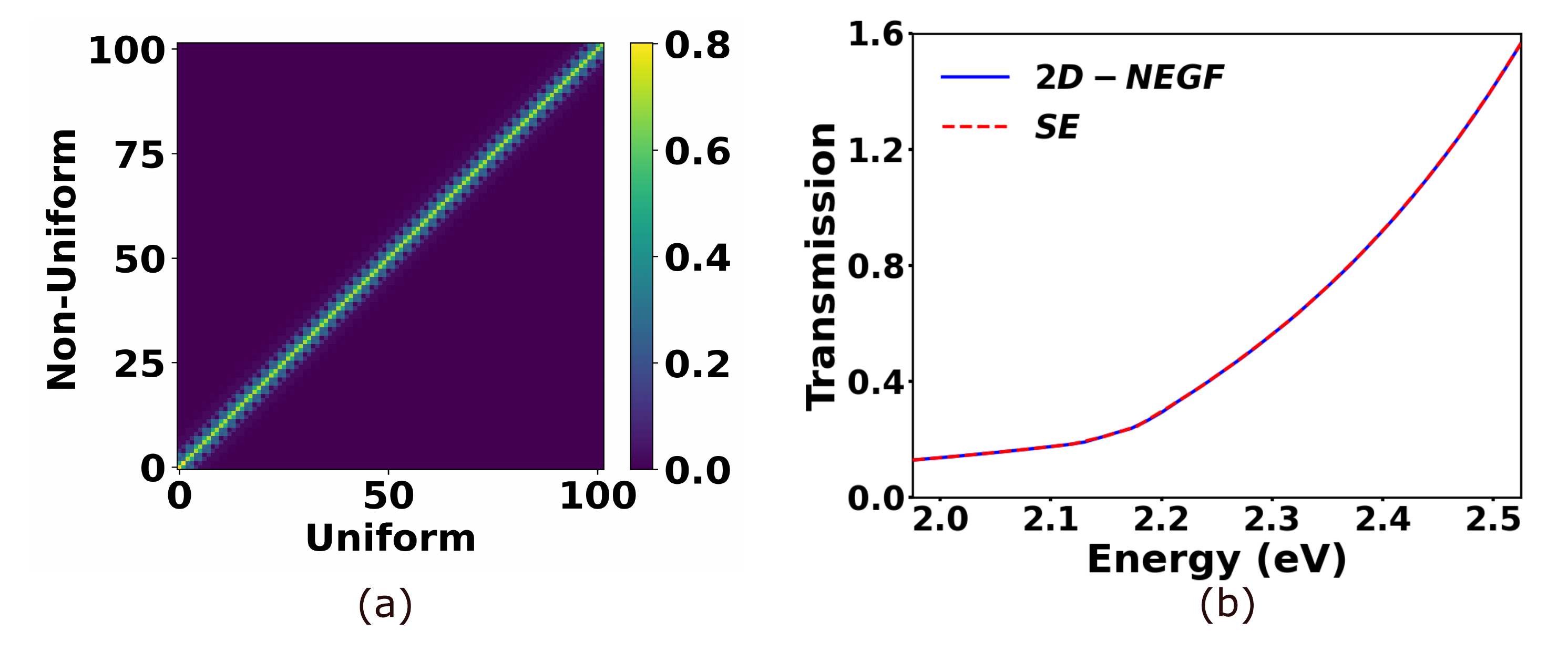}
    \caption{(a) Inner-product of eigenstate of transverse Hamiltonian of uniform and non-uniform contact. (b)  The transmission over different energy level using full 2D-NEGF and spatio-eigen (SE) NEGF with 50 transverse eigen($i$)-channels.}
    \label{fig:Eig2D}
\end{figure}

First, we use the complete 2D-NEGF method to calculate the charge current. The result of this calculation serves as a benchmark. Moving on to the spatio-eigen approach, we use the eigenvectors of the uniform contact Hamiltonian for the transverse direction (\(H^{FM}_B\)) and transform the non-uniform contact's Hamiltonian to that basis $H^{FM}_T$ using a unitary transformation (Eq. \ref{eq:Uij}). Before presenting the simulation results, we first discuss the assumption in Eq.\ref{eq:u12} that each transverse eigen ($i$) channel is connected to only a limited number of channels. A trivial reason is that the sum of squares of rows and columns of a unitary matrix should be one. Additionally, eigen-basis with similar wavelengths exhibit higher overlap. As we move to lower or higher transverse energy eigenvectors while keeping one eigenvalue constant in Eq.\ref{eq:Uij}, the overlap decreases because the wave number increases with energy. In short, transverse eigen ($i$) channels with similar kinetic energy terms have higher overlap. We expect \(\mathbf{U}\) to feature a region of high overlap near the diagonal (as the eigenvalues of both contacts are of the same order). Examining the structure of \(U\) (Fig.~\ref{fig:Eig2D}(a)) for our 2D-MTJ, we find that each eigenvector significantly overlaps with only a small set of eigenvalues, validating our assumption. This assures the applicability of our approach, which is successfully demonstrated in the following paragraph.\\
\begin{figure}
    \centering
    \includegraphics[width=\linewidth]{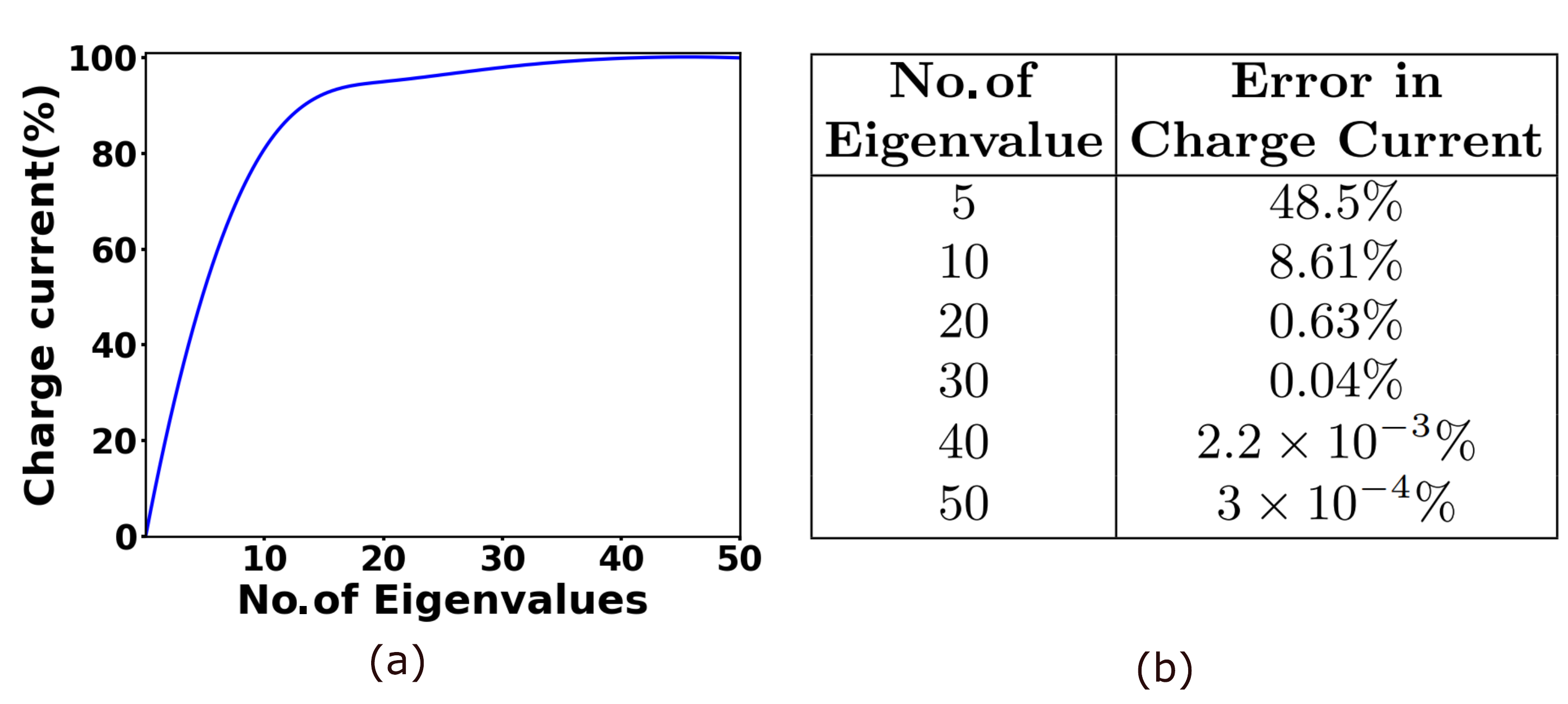}
    \caption{(a) Charge current in the spatio-eigen basis (as a percentage of the charge current in 2D NEGF), varying with the number of relevant eigen-channel. (b) Error in the charge current in the spatio-eigen basis compared to the 2D NEGF.}
    \label{fig:Sheet}
\end{figure}
To observe the diminishing impact of irrelevant eigenvalues, we start with a small relevant set and gradually increase its size. As depicted in Fig. \ref{fig:Sheet}, we observe that the charge current obtained using the spatio-eigen approach converges to the charge current obtained with 2D NEGF as the size of the eigenvalue set increases. This comparison provides valuable insights into the accuracy and efficiency of our computational approach.\\
When the simulation considers 30 eigenvalues, our method yields remarkably accurate results, displaying an error of only 0.04\%. The error decreases further to below $3 \times 10^{-3}\%$ with only 50 eigenvalues. We also observe an error of less than 0.05\% in the spin current for a set of 30 eigenvalues.
Examining the transmission at different energy levels (see Fig.~\ref{fig:Eig2D}(b)), we observe that both the 2D-NEGF and our spatio-eigen approach yield identical transmission across all energy levels. A similar trend is observed for different device characteristics and non-uniform contacts. Hence, this toy model demonstrates that the argument we presented in the methodology section regarding the reduction of the effects of conduction and interference from the irrelevant set of eigenstates holds true as we increase the threshold of the relevant set. The toy model numerically demonstrates that we can capture all the physics of quantum transport using spatio-eigen approaches of NEGF while utilizing only a fraction of computational resources compared to full 3D-NEGF.

\section{Sk-MTJ and eigenvectors overlap}
\label{appB}
\begin{figure}
    \centering
    \includegraphics[width=\linewidth]{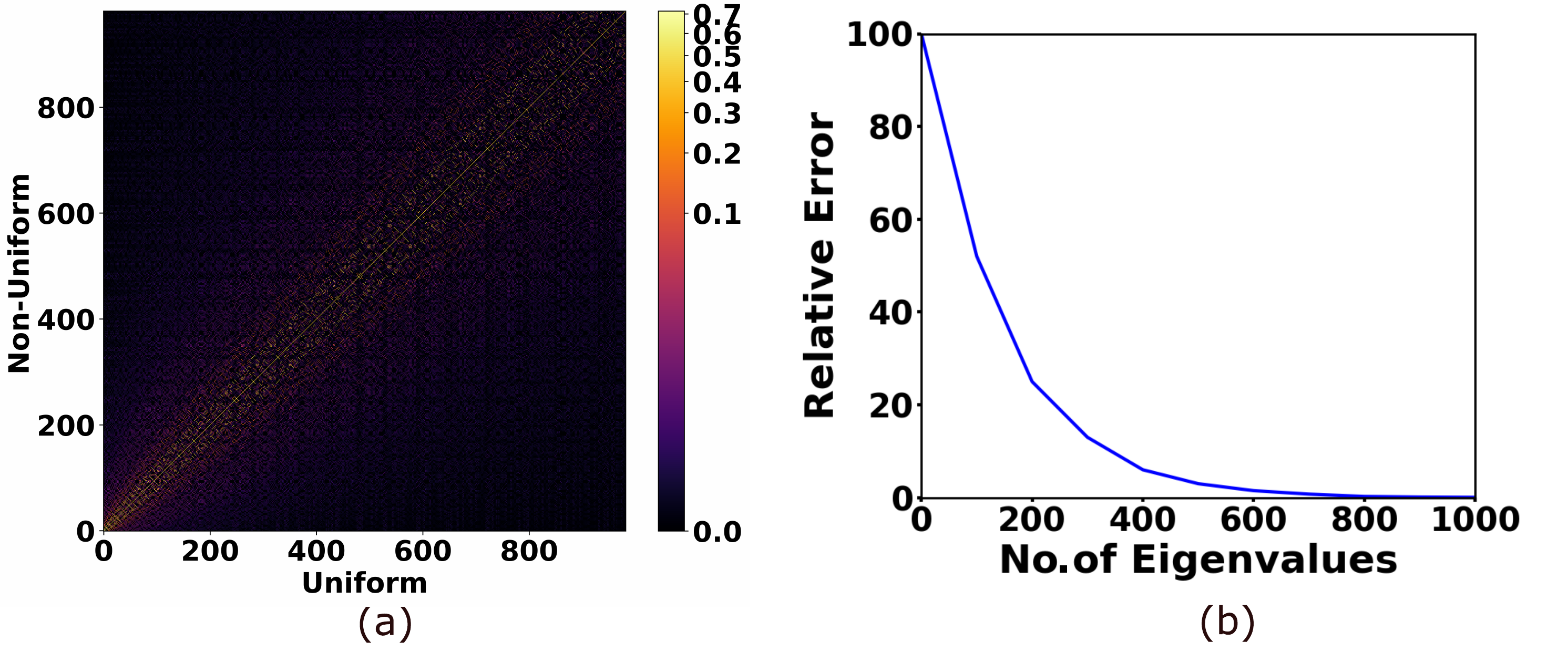}
    \caption{  (a) Inner-product of transverse eigenstate of uniform and non-uniform contact (b) Change in current operator with introduction additional 100 eigen values.}
    \label{fig:Eig3D}
\end{figure}
Building on our discussion of the 2D-MTJ, the 3D structure is not fundamentally different from its 2D counterpart; it simply has one additional degree of freedom. Therefore, we expect our spatio-eigen approach to be applicable to the 3D model as well. To implement the spatio-eigen based NEGF for Sk-MTJ, we have included 900 transverse eigen-channels in our relevant set. Examining the structure of the unitary transformation matrix (Fig. \ref{fig:Eig3D}(a)), we observe that each eigen-channel has significant overlap with only a small set of eigenvalues, further validating our assumption in Eq. \ref{eq:u12} for our device. Additionally, to demonstrate the diminishing effect of higher-order eigen-channels, we begin with a small set of eigenbasis and gradually increase the threshold, calculating the change in the current operator with the introduction of each new set of eigen-channels. As shown in Fig.~\ref{fig:Eig3D}(b), this change diminishes as the threshold increases, indicating that the current operator converges to a constant matrix. This aligns with our claim that the effect of irrelevant terms reduces as we increase the threshold ($E_t$). 
\section{Inverse of matrix in terms of $2\times2$ block}
\label{app:Mat}
For a matrix divided into a \( 2 \times 2 \) block form, the block-wise inverse is given by
\begin{subequations}
\label{eq:inv2}
    \begin{equation}
    \begin{array}{l}
    \left(\begin{array}{ll}
    a & b \\
    c & d
    \end{array}\right)
    =\left(\begin{array}{ll}
    A & B \\
    C & D
    \end{array}\right)^{-1}
    \end{array}
    \end{equation}
    
    \begin{equation}
    a = (A - B D^{-1}C)^{-1}
    \end{equation}
    \begin{equation}
        c = D^{-1}Ca
    \end{equation}
    \begin{equation}
    d = (D - C A^{-1}B)^{-1}
    \end{equation}
    \begin{equation}
        b = A^{-1}Bd
    \end{equation}
\end{subequations}

\nocite{*}
\noindent\textit{Acknowledgment:} The author Abhishek Sharma acknowledges the support by the Science and Engineering Research Board (SERB), Government of India, Grant No. SRG/2023/001327 and the initial support provided by Prof. Subrahmanyam Murala for this project.\\
\noindent\textit{Data Availability:} The data and that support the findings of this study are available within the article. The developed code is also available on Github (\href{https://github.com/devimat/SpatioEigen}{devimat/SpatioEigen}) upon reasonable request to the corresponding author. 
\bibliography{Reference}

\end{document}